\documentclass[final,3p,times]{elsarticle}

\usepackage{amssymb}
\usepackage{lineno}
\usepackage{graphicx}
\usepackage{xcolor}
\usepackage{layouts}
\usepackage{multicol}
\usepackage{threeparttable}
\usepackage{booktabs}
\usepackage{longtable}
\usepackage{hyperref}
\usepackage{amsthm, amsmath}
\usepackage{algorithm}
\usepackage{algpseudocode}
\usepackage{placeins}
\usepackage{amsfonts}
\usepackage{pgfplots}
\pgfplotsset{compat=1.18}
\usepackage{subcaption}
\usepackage{pgfplots}
\usepackage{xcolor}

\newcommand{\beq}{\begin{equation}}
\newcommand{\eeq}{\end{equation}}

\def\ten   #1{\boldsymbol{#1}}
\def\vec   #1{\boldsymbol{#1}}
\newcommand{\newtext}[1]{\textcolor{black}{#1}}

\journal{arXiv}

\graphicspath{{figs//}}

\begin{document}

\begin{frontmatter}

\title{The mechanics of \newtext{the} \textit{Less In More Out} \newtext{artificial heart}: \\modeling fabric-based soft robotic \newtext{devices}}

\author[inst1]{Marin Lauber}
\author[inst2,inst3]{Maziar Arfaee}
\author[inst1]{Mathias Peirlinck\corref{cor1}}
\ead{mplab-me@tudelft.nl}

\affiliation[inst1]{organization={Department of BioMechanical Engineering, Faculty of Mechanical Engineering, Delft~University~of~Technology~(TU~Delft)},
           addressline={Delft}, 
           country={The Netherlands}
}
\affiliation[inst2]{organization={Department of Cardiothoracic Surgery, Thorax~Center, Erasmus~MC},
            addressline={Rotterdam}, 
            country={The~Netherlands}
}
\affiliation[inst3]{organization={Autonomous Matter Department, AMOLF},
            addressline={Amsterdam}, 
            country={The Netherlands}
}

\begin{abstract}
\newtext{
Recently, the \emph{Less In More Out} device, a fluidically actuated soft total artificial heart was proposed.
This device uses arrays of pouch motors to achieve a positive fluidic lever when pneumatically actuated against physiological hemodynamic conditions.
Extensive experimental characterization demonstrated its potential; however, experiments alone cannot resolve the internal mechanical fields that govern device durability and performance.
Here, we develop a computational framework to investigate intrinsic device mechanics, such as stress concentrations, strain paths, and fatigue life, and to explore targeted design modifications that improve durability and efficiency.
We show that our model reproduces the nonlinear deformation and pressure–volume relationships measured experimentally under varying hemodynamic conditions.
Across designs, devices with fewer pouches deliver higher stroke volumes but exhibit up to 50\% higher peak von Mises stresses, which reduces their fatigue life.
Our simulations further identify heat-sealed seams and buckling regions as durability-limiting features.
As a proof of concept, we vary the valve support aspect ratio and relative endocardial-epicardial pouch fabric compliance, reducing the peak von Mises stress by $\sim10$\% while maintaining identical physiological outputs and improving mechanical efficiency.
Overall, our framework enables detailed evaluation of stress hotspots, buckling, and fatigue life, and offers a foundation for optimizing artificial hearts and other fluidically actuated fabric-based soft robotic devices.}

\end{abstract}

%     \includegraphics[width=\textwidth]{figs/graphical_abstract.pdf}

% \begin{highlights}
% \item Computational model of a fabric soft artificial heart benchmarked with experiments
% \item Model resolves deformation, internal stress, buckling, and strain paths in the device
% \item Fewer pouches increase stroke volume but elevate peak stresses and reduce fatigue life
% \item Heat-sealed seams and edge buckling govern durability-limiting stress concentrations
% \item In silico informed design changes reduce peak stress by 10\% without loss of output
% \end{highlights}

\begin{keyword}
soft robotics \sep artificial heart \sep fabric textile actuator \sep fluidic actuation \sep fatigue analysis \sep numerical design optimization
\end{keyword}

\end{frontmatter}


\section{Introduction}
\label{sec:sample1}
%%%%%%%%%%%%%%%%%%%%%%%%%%%%%%%%%%%%%%%
% general soft robotics
Soft robotics is emerging as a compelling alternative to rigid robotics in many areas of science and engineering, offering enhanced adaptability and compliance through elastomers, silicone, or fabrics while relying on comparatively simple actuation mechanisms \cite{Yasa2023AnRobotics, Roche2014AMaterial}.
Among these, fabric-based devices have gained attention for their low weight, durability, and biocompatibility \cite{Fu2022TextilesTrends}.
\newtext{Fabrics offer superior force transmission and tailorable anisotropy, enabling actuators and artificial muscles capable of contraction, extension, bending, and twisting \cite{Arfaee2023ModelingMotors, Belforte2014BellowsMuscle, Connolly2019Sew-freeRobots,Gorissen2017ElasticApplications, Ge2020DesignGloves, Guo2024EncodedRobots, Lobanov2016ModelingBehavior,  Naclerio2020SimpleMuscle}.}

% soft robotics in the medical field
These attributes make fabric-based soft robotic devices promising for biomedical applications \cite{Cianchetti2018BiomedicalRobotics}.
In particular, combining soft robotics with the development of total artificial hearts offers the potential to mitigate complications associated with rigid mechanical designs and actuation methods \cite{Cohrs2017ACirculation, Guex2021IncreasedHeart, Park2022ComputationalFunction, Feng2024ComparativeBehaviors, Zrinscak2025DesignWall, Ueda2025SoftFunctions, Arfaee2025, Arfaee2025AHeart} 
Soft artificial hearts can reduce thrombogenic risk, generate physiological pulsatile flows through fluidic actuation, and adopt organic geometries via scalable manufacturing methods.
Examples include devices with embedded McKibben muscles to reproduce natural ventricular contraction and twisting \cite{Park2022ComputationalFunction, Osouli2025}, and recent designs relying entirely on fabric-based ventricles and pouch-based actuators that can operate under physiological loading in mock circulatory loops \cite{Arfaee2025, Arfaee2025AHeart}.

% modeling and challenges to address
Despite these advances, there is currently no validated and practically usable numerical framework capable of resolving internal stress, buckling, and fatigue-relevant strain fields in fabric-based, fluid-actuated soft total artificial hearts.
Such a framework is required to study intrinsic device mechanics, identify durability-limiting features, and rationally guide design modifications, tasks that remain inaccessible to purely experimental approaches \cite{Armanini2023SoftOverview}.
For the \emph{Less In More Out} (LIMO) soft total artificial heart recently introduced in \cite{Arfaee2025}, experiments quantified global pressure–volume behavior, stroke volume, efficiency, and failure locations, but did not provide access to the internal stress and strain fields governing fatigue life or their sensitivity to design parameters.
In particular, understanding how buckling, tension-line formation, and localized stress concentrations control fatigue life is essential for durable designs, yet difficult to infer from experiments alone.

The soft nature of these devices leads to large, nonlinear responses that are challenging to characterize comprehensively in bench-top tests.
For soft artificial hearts in particular, design choices must be evaluated not only in terms of global pressure–volume behavior, stroke volume, and afterload sensitivity, but also in terms of local stress and fatigue resistance.
Computational models have therefore become an important complement to prototyping, and have been used to optimize soft robotic devices and artificial hearts \cite{Park2022ComputationalFunction, Ge2020DesignGloves, Kaczmarski2025Ultra-fastTrunk, Osouli2025, Guo2024EncodedRobots}, including studies that couple structural models to lumped-parameter circulation models for patient-specific loading conditions \cite{Fedele2017ASurfaces, Peirlinck2018b, Osouli2025}. For fabric-based devices, however, the interplay of large deformations, bending-driven instabilities, and durability-limiting seams makes it difficult to infer internal mechanical fields experimentally and challenging to model robustly. 

Beyond performance, soft artificial hearts operate under millions of loading cycles, and the fatigue resistance of their constituent materials is often low, making fatigue - and thus limiting stress concentrations - a key design constraint \cite{Guex2021IncreasedHeart, Arfaee2025}.
Advanced experimental techniques, such as Digital Image Correlation  \cite{Hebert2023TheReview}, can provide local strain fields for use in strain–life (E–N) fatigue approaches, but are typically limited to selected configurations and loading scenarios.
Numerical models are therefore particularly useful in this context, because they provide the full-field local strain measures required by E–N and related numerical fatigue frameworks. 
Existing methods range from continuum damage mechanics to decoupled approaches using separate constitutive and damage models \cite{Auricchio2016FatigueAnalysis}.
For woven fabrics, however, fatigue prediction remains challenging: varying weave type and fiber level properties strongly influence fatigue life \cite{Lyons1962FatigueLifetimes, Lyons1965LawFibers}, and suitable experimental data to assess damage progression are scarce, which hampers fully calibrated lifetime predictive models.

In this work, we present the numerical modeling of a soft total artificial heart made of woven fabric and actuated by pressurized pouches, replicating the LIMO device recently introduced and experimentally characterized in \cite{Arfaee2025}.
Our first aim is to benchmark 
the model against these static inflation measurements under different pouch numbers and afterloads, and thereby establish a reliable framework for further studies.
Our second aim is to use the model to investigate intrinsic mechanical responses, including stress concentration, buckling, strain directionality, and fatigue risk, in order to rationalize observed failure locations and identify durability-limiting features.
\newtext{Our third aim is to demonstrate how targeted geometric and material modifications, such as changes in valve support geometry and endo-vs-epicardial pouch compliances, can reduce peak stresses while preserving stroke volume and mechanical efficiency.}
We begin by describing the device geometry, material properties, and numerical implementation (Section~\ref{sec:method}), followed by a comparison of simulated and experimental pressure–volume relationships (Section~\ref{sec:results}).
We then use the full-field simulation results to examine local stress concentrations, buckling behavior, and fatigue life\newtext{, and to analyze the impact of selected design modifications}.
Finally, in Section~\ref{sec:discussion} we discuss the implications of these findings, outline our limitations, and suggest directions for future work.

%%%%%%%%%%%%%%%%%%%%%%%%%%%%%%%%
\section{Methods}
\label{sec:method}
%%%%%%%%%%%%%%%%%%%%%%%%%%%%%%%

%%%%%%%%%%%%%%%%%%%%%%%%%%%%%%%%
\subsection{Geometry}
\label{sec:method_geom}
%%%%%%%%%%%%%%%%%%%%%%%%%%%%%%%%
\textbf{Real device}.
The physical LIMO device is an airtight fabric ventricle \cite{Arfaee2025} made of thermoplastic polyurethane (TPU)-coated nylon fabric, approximately 0.2 mm thick (Riverseal 70 LW, 78 Dtex; 170 g/m²; Rivertex, Culemborg, The Netherlands).
It is about 140 mm in length and holds 300 ml when fully inflated.
As shown in Figure~\ref{fig:kinematics}, the device features flat pouches on its surface that can be fluidically actuated, i.e. inflated into roughly cylindrical shapes (Figure~\ref{fig:kinematics}).
These pouches consist of two layers of heat-sealed fabric, which are then folded along one edge and sealed onto themselves to form the ventricle.
This process produces an odd number of pouches, with one pouch located along the longitudinal fold line.
The pouches are inflated by compressed air, reducing the internal volume of the ventricle and ejecting fluid, thereby mimicking the natural contraction of a healthy human heart \cite{Peirlinck2021}.
The ventricle attaches to a rigid valve support providing two valves for single ventricle function.
Two such ventricles combined form a complete bi-ventricular soft artificial heart.
During the experimental characterization, a water column is connected to the device to provide a fixed ventricular pressure $P_v$.
The device is actuated through a pressure-controlled valve, inflating the pouches. 
\newtext{In the experimental protocols, the pouch pressure is increased at a rate of 0.25 kPa/s, which is sufficiently slow to neglect inertial fluid effects.}

\textbf{Numerical idealization}.
Our computational model uses a simplified, symmetric geometry with an even number of pouches on the front and back surfaces.
This differs from the physical device, which has an odd number of pouches.
\newtext{This odd fold-line pouch is subject to strong buckling and folding, which limits its effectiveness, see Figure~4B of the original manuscript. 
Additionally, self-contact occurs within this folded pouch during buckling and folding and would require explicit contact modeling, resulting in prohibitively high computational costs.
For these reasons, we omit this odd fold-line pouch and focus on front/back symmetric devices in this study.}
The physical device has a single pressure inlet and uses discontinuous seams to create inter-pouch connections, distributing the actuation pressure evenly among them.
We omit these discontinuities in our numerical model, and seams run continuously along the pouches' length. This simplification preserves the dominant load paths and stress localizations relevant for our durability assessments.
Lastly, we model the valve support without valves, as they are not required to replicate the inflation scenarios considered in the original experimental study.

%%%%%%%%%%%%%%%%%%%%%%%%%%%%%%%%
\subsection{Kinematics}
\label{sec:kinematics}
\begin{figure}
    \includegraphics[width=\textwidth]{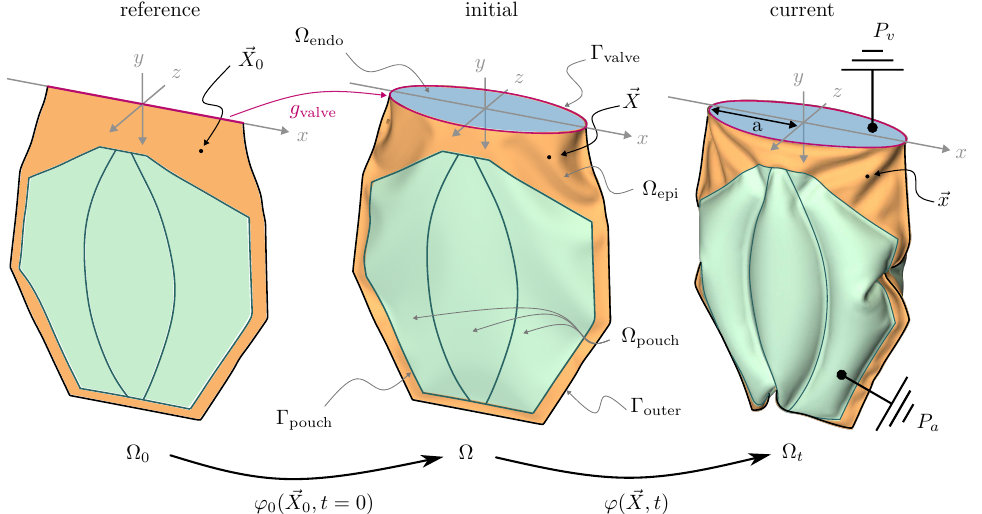}
    \caption{\textbf{Computational domain of the soft ventricle and deformation during inflation.} Isometric view of the computational domain $\Omega_0=\Omega_{\text{endo},0} \cup \Omega_{\text{epi},0} \cup \Omega_{\text{pouch},0}$ with the valve boundary $\Gamma_{\text{valve},0}$ at different times during the inflation. (left) Reference configuration of the device before being fitted to the valve support. (center) Initial configuration of the device after the fitting step.
    (right) Current configuration at the end of the inflation with a static pressure $P_v=13.8$ kPa.}
    \label{fig:kinematics}
\end{figure}

%%%%%%%%%%%%%%%%%%%%%%%%%%%%%%%%
The geometry of the device undergoes two major geometric transformations as shown in Figure~\ref{fig:kinematics}:
first, it transitions from the flat reference geometry to the initial pre-loaded ventricular and pre-inflated pouch configuration fitted onto the \newtext{rigid} valve support; second, it transitions from this initial state to the loaded and inflated configuration.
The first-stage mapping to the initial configuration accounts for geometric changes and pre-strain introduced during attachment to the valve support and is explicitly resolved prior to the inflation analysis.  

\textbf{Mapping definitions}.
% \mpnote{@Marin, do we still need the time t notes here?}
The reference configuration $\Omega_0\subset\mathbb{R}^3$ corresponds to the flat state of the device before it is fitted to the valve support.
The initial configuration $\Omega$ is obtained through the mapping $\varphi_0$,
\begin{equation}
    \varphi_0(\vec{X}_0,t) : \Omega_0\times\mathbb{R}\to\mathbb{R}^3 
    \quad \text{such that}\quad 
    \vec{X}=\varphi_0(\vec{X}_0,t=0) ,
\end{equation}
where $\vec{X}_0$ is a point in $\Omega_0$ and $\vec{X}$ is the corresponding point in $\Omega$.  
The current (inflated) configuration $\Omega_t$ is obtained from $\Omega$ via $\varphi$:  
\begin{equation}
    \varphi(\vec{X},t) : \Omega\times\mathbb{R}\to\mathbb{R}^3 
    \quad \text{such that}\quad 
    \vec{x}=\varphi(\vec{X},t) ,
\end{equation}
where $\vec{x}$ is a position in $\Omega_t$.  
Mapping directly from $\Omega_0$ to $\Omega_t$ is given by $\hat{\varphi}$,  
\begin{equation}
    \hat{\varphi}(\vec{X}_0,t): \Omega_0\times\mathbb{R}\to\mathbb{R}^3 
    \quad \text{such that}\quad 
    \vec{x}=\hat{\varphi}(\vec{X}_0 ,t) .
\end{equation}

\textbf{Deformation gradient.}  
The total deformation from $\Omega_0$ to $\Omega_t$ is described by $\hat{\varphi}(\vec{X}_0,t)$.  
The associated total deformation gradient is
\begin{equation}
    \mathbf{F}(\vec{X}_0,t) = \frac{\partial \hat{\varphi}(\vec{X}_0,t)}{\partial \vec{X}_0}.
\label{eq:totdefgrad}
\end{equation}
This total deformation gradient can be multiplicatively decomposed into contributions from the fitting step (mapping $\varphi_0$) and the inflation step (mapping $\varphi$):
\begin{equation}
    \mathbf{F} = \mathbf{F}_{\varphi} \, \mathbf{F}_{\varphi_0}, 
    \quad
    \mathbf{F}_{\varphi} = \frac{\partial \varphi(\vec{X},t)}{\partial \vec{X}}, 
    \quad
    \mathbf{F}_{\varphi_0} = \frac{\partial \varphi_0(\vec{X}_0)}{\partial \vec{X}_0}.
\end{equation}
Here $\mathbf{F}_{\varphi_0}$ represents the pre-stretch induced during the fitting of the membrane to the valve support, while $\mathbf{F}_{\varphi}$ describes the subsequent deformation due to inflation.

\textbf{Geometric parametrization.}  
We divide the computational domain $\Omega_0$ into the endocardium $\Omega_{\text{endo},0}$, the epicardium $\Omega_{\text{epi},0}$, and the pouches $\Omega_{\text{pouch},0}$.  
The edges of these subdomains are $\Gamma_{\text{valve},0}$ for the valve boundary, $\Gamma_{\text{pouch},0}$ for the pouch seams, and $\Gamma_{\text{outer},0}$ for the outer edge.  

The outer edge curves $\{\Gamma_{\text{outer},0},\,\Gamma_{\text{valve},0} \}$ and the seam lines of the pouches $\Gamma_{\text{pouch},0}$ fully parameterize the geometry, see Figure~\ref{fig:kinematics}.  
We then fit planar patches through these curves and obtain the different surfaces of the model.

\textbf{Computation of the initial configuration.}  
We replicate the physical construction process where the flat stack of fabric is fitted to a \newtext{rigid} valve support with an ellipsoidal cross-section.
The initial map $\varphi_0$ from the flat state of the device $\Omega_0$ onto this 3D geometry (to $\Omega$ in Figure~\ref{fig:kinematics}) induces pre-stretch in the fabric and initializes some of the folds present in the device.
To compute this configuration, we treat the flat reference geometry $\Omega_0$ as the starting point and apply displacement constraints to match the reference outer edge $\Gamma_{\text{valve},0}$ to the initial valve support $\Gamma_{\text{valve}}$.  
The \newtext{original LIMO design} valve support has an aspect ratio $\Lambda=2$ and a semi-major axis length ratio $a/L\sim0.3275$, where $L$ is the device length. As such, all nodes on $\Gamma_{\text{valve}}$ in the current configuration satisfy
\begin{equation}
    g_\text{valve}(\vec{x}) =  
    \begin{cases} 
        u, w \quad\big| (X+u)^2 + \frac{1}{\Lambda^2} (Z+w)^2 = a^2, \\ 
        v = 0 
    \end{cases} 
    \quad \forall \vec{x} \in \Gamma_\text{valve},
\label{eq:gvalve}
\end{equation}
where $(X,Y,Z)$ are nodal reference coordinates on $\Gamma_{\text{valve},0}$ in the reference configuration and $(u,v,w)$ represents the Cartesian displacement vector to map these nodes onto $\Gamma_{\text{valve}}$ in the current configuration.  

To avoid introducing artificial local stretching of the LIMO along $\Gamma_\text{valve}$ during this computational mapping procedure, we impose the additional constraint
\begin{equation}
    \Delta l_0 = \Delta l = \oint_{0}^{l}C(\tau)\, \mathrm{d}\tau \quad \forall\, l \in (0,\pi],
\label{eq:gvalvestretch}
\end{equation}
where $l$ is an angular coordinate along $\Gamma_\text{valve}$, $C(\tau)$ denotes the local line element arc length along the valve edge, $\Delta l_0$ and $\Delta l$ denote the segment lengths in the reference and current configurations. 
Together with the displacement constraint in Eq.~\ref{eq:gvalve}, this inextensibility condition keeps the valve edge stretch-free while allowing the remainder of the fabric to deform to fit the support, thereby inducing pre-stretch in the pouches and walls.
\newtext{We apply these constraints progressively by linearly ramping the prescribed displacements $(u,v,w)$ during the initial stage of our simulation.} 
Figure~\ref{fig:kinematics} shows the device before (left) and after (center) the computation of this initial configuration.

%%%%%%%%%%%%%%%%%%%%%%%%%%%%%%%%
\subsection{Governing equations}
\label{sec:equations}
%%%%%%%%%%%%%%%%%%%%%%%%%%%%%%%%

The mechanical behavior of the soft fabric ventricle is governed by the balance of linear momentum
\begin{equation}\label{eq:dynamic}
    \rho \frac{\text{D}\Vec{u}}{\text{D} t} - \nabla\cdot{\Vec \sigma} - {\Vec f} = 0 \quad \forall \Vec{x} \, \in \Omega \times (0,T],
\end{equation}
where $\vec{u} \in \mathbb{R}^3\equiv u\hat{\vec{x}}+v\hat{\vec{y}}+w\hat{\vec{z}}$, $\ten{\sigma}\in \mathbb{R}^{3\times3}$ and $\Vec{f}\in \mathbb{R}^3$ are the displacement vector, Cauchy stress tensor, and external force vector, respectively. 
The Cartesian basis vectors are denoted $\hat{\vec{x}}, \hat{\vec{y}}, \hat{\vec{z}}$.
Eq.~\ref{eq:dynamic} states that, at each material point, the inertial forces, internal stresses, and external forces must be in equilibrium.

The second Piola-Kirchhoff stress tensor, $\ten{S}\in \mathbb{R}^{3\times3}$, can be obtained from a pull-back operation on the Cauchy stress tensor, 
\begin{equation}
    \ten{S} = J\ten{F}^{-1}{\Vec \sigma}\ten{F}^{-\top}
\end{equation}
where $\ten{F}\in \mathbb{R}^{3\times3}$ is the total deformation gradient tensor (Eq. \ref{eq:totdefgrad}) and $J\equiv\det\ten{F}\in \mathbb{R}$.
To ensure thermodynamic consistency, we introduce the Helmholtz free energy as a function of the deformation gradient $\psi(\ten{F})$.
Assuming no dissipative energy losses within the material, and rewriting the Clausius-Duhem entropy inequality following the Coleman and Noll principle \cite{Peirlinck2024c}, we derive the second Piola-Kirchhoff stress tensor
\begin{equation}
    \ten{S}=\frac{\partial\psi(\ten{F})}{\partial \ten{E}}
\end{equation}
with the Green-Lagrange strain tensor $\ten{E} = \frac{1}{2}\left(\ten{C}-\mathbb{I}\right)\in \mathbb{R}^{3\times 3}$ and the right Cauchy-Green deformation tensor $\ten{C} = \ten{F}^{\top}\ten{F}\in \mathbb{R}^{3\times 3}$. Eq. \ref{eq:dynamic} is subject to the initial conditions
\begin{equation}
    {\Vec u}({\Vec x},0) = {\Vec 0}, \quad \dot{\Vec u}({\Vec x},0) = {\Vec 0} \quad \forall \vec{x} \in \Omega
\end{equation}
and the Dirichlet initial boundary conditions on the valve support
\begin{equation}
    {\Vec u}({\Vec x},t) = g_\text{valve}({\Vec{x}}) \quad \forall \Vec{x} \in \Gamma_\text{valve}.
\end{equation}
as discussed in Eqs. \ref{eq:gvalve} and \ref{eq:gvalvestretch}.
\newtext{In the original experiments, LIMO prototypes were tested against different afterloads using a water column connected to the ventricle. In the simulations, this internal fluid loading is represented by a uniform normal pressure $P_v$ applied to the endocardial surface, i.e. by a Neumann boundary condition,}
\begin{equation}\label{eq:stress_BC}
    {\Vec F}{\Vec S}\hat{\Vec n} = -P_vJ{\Vec F}^{-\top}\hat{\Vec n} \quad \forall \Vec{x} \in \Omega_{\text{endo}} \times (0,T],
\end{equation}
where $\hat{\vec n}\in \mathbb{R}^3$ is the unit normal vector and $P_v\in \mathbb{R}$ is the ventricular pressure.
\newtext{In the experiment of \cite{Arfaee2025}, this ventricular pressure is the sum of an initial ventricular pressure $P_{v,0}$ and an afterload pressure $P_\text{aft}$. 
To test the device under different hemodynamic conditions, they vary the afterload pressure ($P_\text{aft}$), resulting in changes in ventricular pressure $P_v$.
For simplicity, in the remainder of this paper, we only refer to the total ventricular pressure $P_v$, but it is understood that when $P_v$ varies, it corresponds to an increase in the afterload in the experiments.
Similarly, the actuation pressure inside the pouches is modeled as a uniform pressure,
\begin{equation}\label{eq:stress_BC_actuation}
    {\Vec F}{\Vec S}\hat{\Vec n} = -P_a(t)J{\Vec F}^{-\top}\hat{\Vec n} \quad \forall \Vec{x} \in \Omega_{\text{pouch}} \times (0,T],
\end{equation}
where $P_a(t)\in \mathbb{R}$ is the prescribed time-dependent actuation pressure.}
In this work, \newtext{we replicate the experimentally enforced pressure loading rate of 0.25 kPa/s by prescribing the same actuation pressure $P_a(t)$ rate, while the ventricular pressure $P_v$ is held constant.} 
The epicardium $\Omega_{\text{epi}}$ is left traction-free in the present inflation simulations.

%%%%%%%%%%%%%%%%%%%%%%%%%%%%%%%%
\subsection{Constitutive modeling}
\label{sec:material_model}
%%%%%%%%%%%%%%%%%%%%%%%%%%%%%%%%
\newtext{Accurate, fabric-specific constitutive models for woven TPU-coated nylon generally require multi-axial tensile data for calibration \cite{McCulloch2024AutomatedFabrics, Peirlinck2024a}, which are not available for the material used in the LIMO heart.}
In the present work, we therefore adopt a homogeneous, isotropic St. Venant–Kirchhoff solid to represent the coated fabric and focus on capturing overall deformation patterns and stress distributions rather than detailed anisotropic effects.
Given the relatively stiff, weakly compliant character of the coated fabric, this choice is appropriate for modeling large geometric displacements combined with relatively small local strains.
For this St. Venant–Kirchhoff material model, the second Piola-Kirchhoff stress tensor is given by
\begin{equation}
    \ten S = \frac{\partial \psi}{\partial \ten E} = \frac{\partial}{\partial\ten{E}}\left(\frac{\lambda}{2}\left[\,\text{tr}\left(\ten{E}\right)\right]^2 + \mu\,\text{tr}\left(\ten{E}^2\right)\right) = \lambda\,\text{tr}\left(\ten E\right)\,\mathbb{I} + 2\mu\ten E,
\label{eq:stvenantkirchoff}
\end{equation}
where $\lambda$ and $\mu$ are the Lamé constants.
\newtext{To parameterize this isotropic law, we use uniaxial tensile test data of the used TPU-coated nylon sheets to identify an in-plane Young’s modulus.
We fit the linear part of the uniaxial stress–strain response and obtain an effective Young’s modulus $E = 0.267$ GPa (see \ref{appendix:calibration}).
The Poisson’s ratio is taken as $\nu = 0.33$, a representative value for slightly compressible polymeric materials in the absence of reliable transverse strain measurements for the coated fabric.
This yields $\lambda = 0.195$ GPa and $\mu = 0.101$ GPa.
This effective linear elastic law in Green–Lagrange strain provides a first-order approximation of the in-plane stiffness and stress distribution for the present analysis.}

%%%%%%%%%%%%%%%%%%%%%%%%%%%%%%%%
\subsection{Numerical solver}
\label{sec:solver}
%%%%%%%%%%%%%%%%%%%%%%%%%%%%%%%%
We use the open-source finite element software \emph{CalculiX} \cite{Dhondt2004} to solve the governing equations of the problem.
\newtext{The balance of linear momentum in Eq.~\ref{eq:dynamic} is integrated using an implicit dynamic solution procedure, replicating the experimentally applied pressure loading rates as our numerical loading ramps.
To avoid spurious oscillations of the model, we use the Hilber, Hughes, and Taylor (HHT) integration method \cite{Hilber1977ImprovedDynamics} with parameter $\alpha=-1/3$ and additional mass-proportional damping taken as 100 times the mass matrix.}
We discretize our geometries with quadratic hexahedral continuum shell elements (S8), resulting in meshes of approximately $10^4$ elements, depending on the specific geometry, and about $2.5\times 10^5$ degrees of freedom.
\ref{appendix:solver_validation} and \ref{appendix:mesh_convergence} demonstrate that this element type accurately captures complex nonlinear deformation and buckling under internal pressure.  
The resulting discretization error is approximately 2\%, which we consider sufficient for the analyses in this work.
True membranes exhibit negligible bending stiffness, leading to an asymptotically infinite number of wrinkles and folds \cite{Cerda2003}.
As we expect our TPU-coated nylon fabric to have some intrinsic finite bending stiffness,  we approximate this fabric as a thin, stiff shell. This modeling choice concomitantly suppresses wrinkles with characteristic lengths smaller than the element size while preserving the global deformation patterns of the underlying structure.

%%%%%%%%%%%%%%%%%%%%%%%%%%%%%%%%
\subsection{Cavity volume calculation}
\label{sec:cavity_volume}
%%%%%%%%%%%%%%%%%%%%%%%%%%%%%%%%
Following our numerical solution, we compute the resulting ventricular volume ($V_v$) and total, summed pouch volume ($V_a$) as a post-processing step using the divergence (Gauss) theorem:
\begin{equation}
    V_\phi({\vec u},t) = \frac{1}{3}\int_{\Omega_{t,\phi}}{\vec x}\cdot\hat{\vec n}_{\text{out}}\, q_\phi \text{ d}S=\frac{1}{3}\int_{\Omega_{t,\phi}}\left({\vec X}+\vec{u}\right)\cdot\hat{\vec n}_{\text{out}}\, q_\phi \text{ d}S
\end{equation}
where $q_\phi$ equals unity for surfaces representing the outer boundary of volume $\phi$ and zero for others. $\hat{\vec n}_{\text{out}}$ denotes the outward-facing unit normal vector of volume $\phi$ and $\Omega_{t,\phi}$ denotes the closed surface enclosing volume $\phi$.
This integration is carried out in the current configuration by first converting the quadrilateral mesh into a triangular mesh and using a single integration point per element.
The method is first-order accurate and provides a fast post-processing estimate.

%%%%%%%%%%%%%%%%%%%%%%%%%%%%%%%%
\section{Results}
\label{sec:results}
%%%%%%%%%%%%%%%%%%%%%%%%%%%%%%%%
In this section, we present the results of our numerical simulations of soft LIMO devices subjected to various experimentally tested ventricular pressures \cite{Arfaee2025} (through varying afterloads).
We first analyze the deformation patterns and the pressure–volume ($P$–$V$) relationship during pouch inflation, followed by an investigation of buckling events, afterload sensitivity, mechanical efficiency, local stress concentrations, fatigue risk, and principal stretch fields.
We then investigate the effect of altering the geometry of the valve support on the stroke volume and local stress measures.

%%%%%%%%%%%%%%%%%%%%%%%%%%%%%%%%
\subsection{Deformation and pressure-volume relationship}
\label{sec:pressure_volume}
%%%%%%%%%%%%%%%%%%%%%%%%%%%%%%%%

\begin{figure}[!ht]
    \centering
    \includegraphics[width=\textwidth]{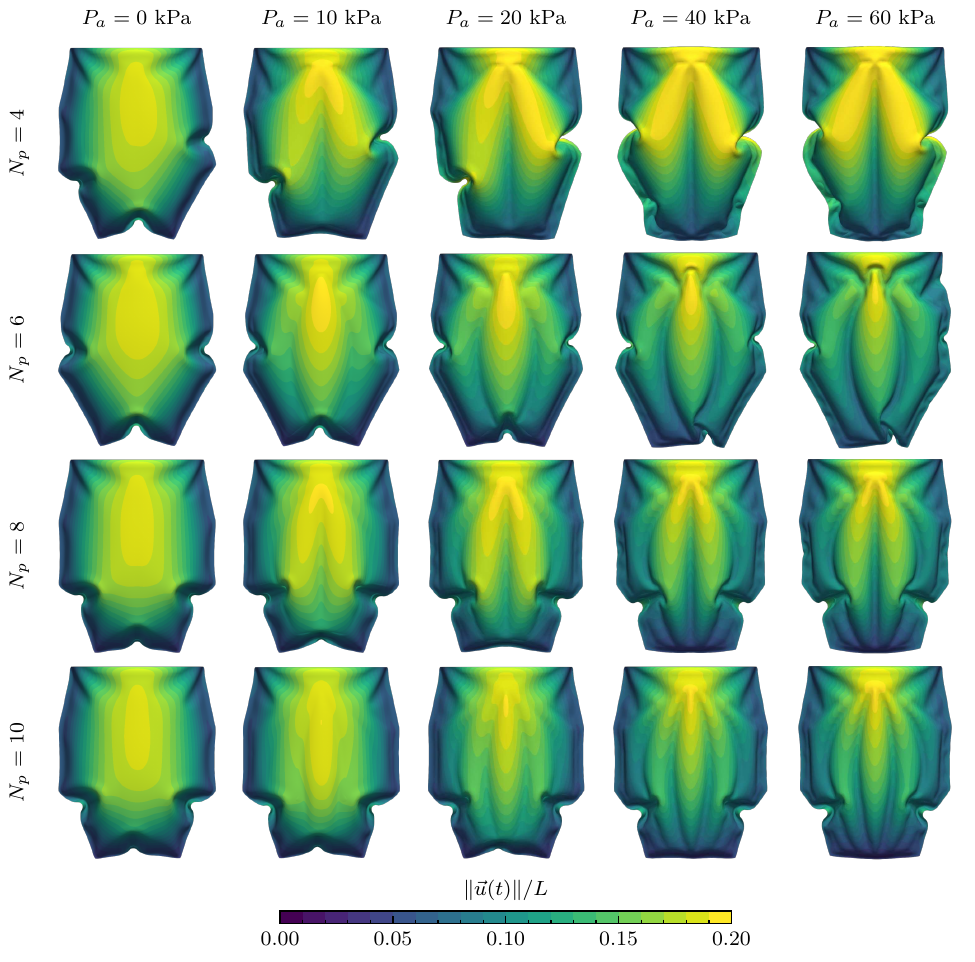}
    \caption{\textbf{Deformation during pouch inflation under \textit{zero-afterload} conditions.} 
    Devices with pouch numbers $N_p \in \{4,6,8,10\}$ are shown under a ventricular pressure of $P_v = 3.8$ kPa, corresponding to the experimental \textit{zero-afterload} condition. Rows correspond to increasing pouch number, and columns correspond to increasing actuation pressure during inflation. The leftmost column shows the initial configuration ($P_a = 0$ kPa), while the rightmost column shows a configuration close to the maximum actuation pressure ($P_a = 60$ kPa), at which the stroke volume is near its maximum.}
    \label{fig:deformation_aft0}
\end{figure}

We first consider devices with pouch numbers $N_p \in \{4,6,8,10\}$ under a ventricular pressure $P_v = 3.8$ kPa ($\sim$28.5 mmHg), corresponding to the experimental \textit{zero-afterload} condition \cite{Arfaee2025}.
Starting from the initial configuration shown in Figure~\ref{fig:kinematics}, the pouches are initially unpressurized ($P_a = 0.0$ kPa) and the ventricle deforms under the applied ventricular pressure $P_v$.
The actuation pressure $P_a$ is then increased linearly at a rate of 0.25 kPa/s up to its maximum values, driving pouch inflation and ventricular contraction.
Figure~\ref{fig:deformation_aft0} shows the resultation deformation at five representative actuation pressures.
As $P_a$ increases, the pouches inflate, the device reconfigures, and the ventricular cavity progressively contracts, leading to a reduction in internal volume from left to right.

\begin{figure}[!ht]
    \centering
        \includegraphics[width=\textwidth]{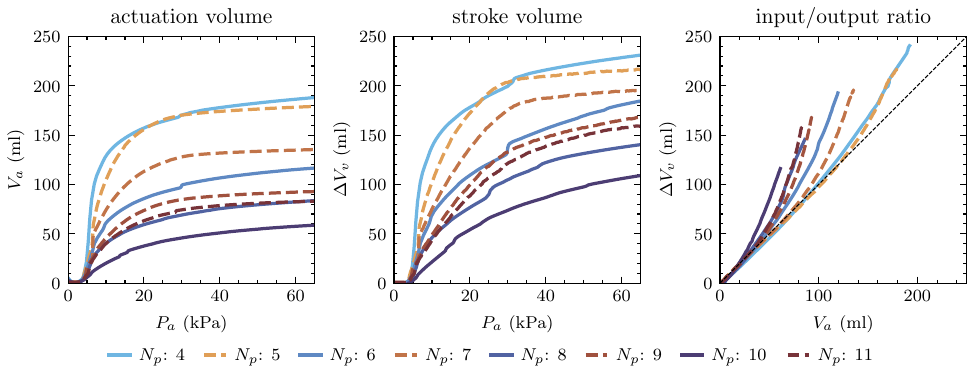}
        \caption{\textbf{Pressure-volume relationship of the soft ventricle under a ventricular pressure of $3.8$ kPa for various pouch numbers $N_p\in\{4,6,8,10\}$}; (Left) change in actuator (pouch) volume against the prescribed actuation pressure; (center) change in the stroke volume against the prescribed actuation pressure and,
        (right) input/output ratio of the soft ventricle with experimental data for $N_p\in\{5,7,9,11\}$ from \cite{Arfaee2025}.}
    \label{fig:pressure_vs_volume}
\end{figure}

The resulting pressure–volume relationships predicted by our LIMO prototypes numerical models are presented in Figure~\ref{fig:pressure_vs_volume}.
At low actuation pressure, pouch pressurization maintains the actuation volume $V_a$, i.e. the \textit{In} in \textit{LIMO}, close to zero (Figure~\ref{fig:pressure_vs_volume} - left), as pouch inflation does not occur until the actuation pressure exceeds the ventricular pressure.
Once $P_a \approx P_v$, the pouches inflate and the device rapidly reconfigures into its contracted state, leading to a sharp increase in $V_a$, which occurs at approximately $P_a \simeq 20$ kPa. At higher actuation pressures, the actuation volume increases more gradually, with the transition between reconfiguration and inflation being most pronounced for devices with fewer pouches.
The ventricular stroke volume $\Delta V_v \equiv V_v(P_a) - V_v(P_a=0)$, i.e. the \textit{Out} in \textit{LIMO}, exhibits a similar response (Figure~\ref{fig:pressure_vs_volume} - center).
It remains small for $P_a < P_v$, then rises rapidly as pouch inflation and global device contraction occur, before increasing more slowly beyond $P_a \sim 20$ kPa. 
As a result, the input–output ratio is initially close to unity (dashed gray lines) but exceeds one for all pouch numbers $N_p$ once global contraction contributes to volume ejection (Figure~\ref{fig:pressure_vs_volume} - right), demonstrating a positive fluidic lever and thus \textit{Less In More Out}.
Experimental measurements from \cite{Arfaee2025} are included in Figure~\ref{fig:pressure_vs_volume} for comparison.
While experiments were conducted on devices with odd pouch numbers, our simulations consider even pouch numbers $N_p \in \{4,6,8,10\}$.
\newtext{To quantify the agreement, experimental $\Delta V_v/V_a$ curves are linearly interpolated to the simulated pouch numbers and compared using a relative $L_2$–norm error. The resulting errors are 7.9\%, 16.8\%, 48.3\%, and 80.3\% for $N_p = 4, 6, 8,$ and $10$, respectively. Agreement is strongest for lower pouch numbers, where deformation is dominated by in-plane stretching, and degrades as pouch number increases and out-of-plane bending becomes more prominent, consistent with the in-plane calibration of the material model.}

%%%%%%%%%%%%%%%%%%%%%%%%%%%%%%%%
\subsection{Buckling and folding}
\label{sec:edge_buckling}
%%%%%%%%%%%%%%%%%%%%%%%%%%%%%%%%
\begin{figure}
    \centering
    \includegraphics[width=\textwidth]{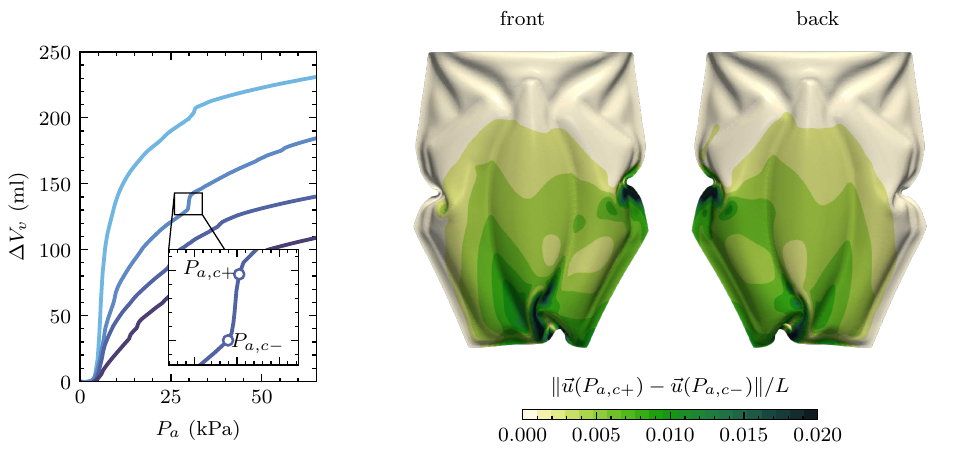}
    \caption{\textbf{Local buckling-induced jump in the pressure-volume response.} 
    Detail of the pressure-volume relationship curve for a representative device with $N_p=6$ under a ventricular pressure of $P_v=3.8$ kPa, highlighting a buckling event occurring at a critical actuation pressure $P_{a,c\pm}$.
    The left panel shows a magnified view of the pressure-volume response around the instability.
    The right panel shows front and back views of the deformed device at $P_{a,c+}$ colored by iso-contours of the normalized displacement difference $\Vert\Vec{u}(P_{a,c+})-\Vec{u}(P_{a,c-})\Vert/L$ across this buckling transition.}
    \label{fig:buckling_close_up}
\end{figure}

Local buckling, i.e. folding, events are observed in most $N_p$ pouch-number configurations and manifest as jumps in stroke volume with increasing actuation pressures in Figure~\ref{fig:pressure_vs_volume} - center.  
To illustrate this phenomenon in more detail, we focus on a representative case with $N_p=6$ under a ventricular pressure $P_v = 3.8$ kPa.
In this case, a step-like change in stroke volume occurs at an approximate actuation pressure $P_a \sim 30$ kPa, associated with localized seam buckling of the seams and sudden reconfiguration of the internal pouch surfaces.
These events produce abrupt changes in the enclosed ventricular volume.
Because the ventricular load is prescribed via a Neumann boundary condition (Eq.~\ref{eq:stress_BC}), fluid inertia does not provide stabilized damping, allowing rapid global reconfigurations to occur.
Figure~\ref{fig:buckling_close_up} highlights this instability: a small pressure increment across the critical actuation pressure $P_{a,c}$ produces a pronounced displacement jump, visualized by iso-contours of the normalized displacement difference between the pre- and post-buckling states. Although abrupt, these instabilities evolve over time scales that remain well resolved by the nonlinear solver time stepping.

\clearpage
%%%%%%%%%%%%%%%%%%%%%%%%%%%%%%%%
\subsection{Afterload sensitivity}
\label{sec:afterload}
\begin{figure}[!ht]
    \centering
        \includegraphics[width=\textwidth]{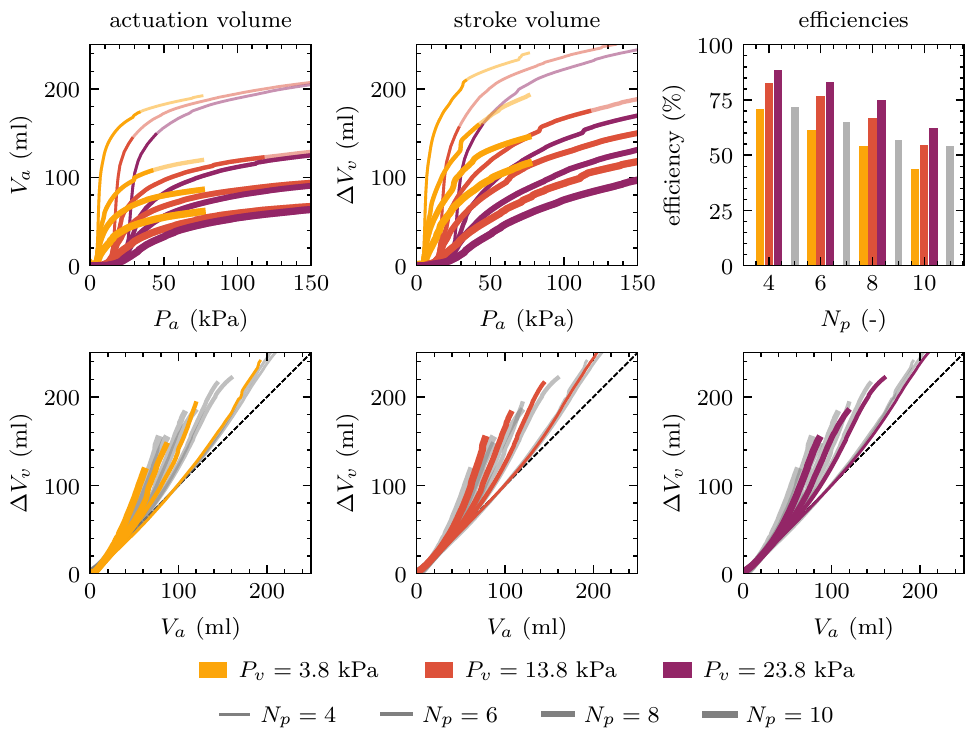}
        \caption{\textbf{Afterload sensitivity and mechanical efficiency across pouch numbers.}
        Top left: Actuation (pouch) volume versus actuation pressure for ventricular pressures $P_v \in \{3.8, 13.8, 23.8\}$ kPa. Line thickness increases with pouch number $N_p$.
        Top center: Ventricular stroke volume versus actuation pressure for the same conditions.
        Top right: Mechanical efficiency evaluated at a physiological stroke volume of $\Delta V_v = 90$ ml. Gray bars indicate experimentally measured efficiencies for $N_p \in \{5,7,9,11\}$ at $P_v = 3.8$ kPa \cite{Arfaee2025}.
        Bottom row: Input–output ratio ($\Delta V_v / V_a$) versus actuation pressure for increasing ventricular pressure from left to right. Light gray curves show responses at different ventricular pressures for comparison.
        }
    \label{fig:afterload_all}
\end{figure}

%%%%%%%%%%%%%%%%%%%%%%%%%%%%%%%%
An important performance aspect of soft total artificial hearts is their ability to operate robustly under varying hemodynamic loading conditions \cite{Guelcher2023}.
Devices with a high afterload sensitivity exhibit a rapid reduction in stroke volume as ventricular pressure and afterload increases, whereas an ideal soft TAH, particularly the left ventricle, should maintain stroke volume and cardiac output across a large range of afterloads.
Here, we examine how changes in ventricular pressure affect the pressure-volume response and mechanical efficiency of the device. 
We vary the ventricular pressure over $P_v \in \{3.8,13.8,23.8\}$ kPa, while applying the same actuation pressure ramp described in Section~\ref{sec:pressure_volume}.
Results for all pouch numbers $N_p$ and ventricular pressures $P_v$ are presented in Figure~\ref{fig:afterload_all}. 
For a fixed pouch number $N_p$ (varying line thickness represents increasing numbers of pouches), increasing $P_v$ (ranging from yellow, to red, to purple for 3.8, 13.8, and 23.8 kPa, respectively) shifts the curves to higher actuation pressures, reflecting the higher actuation pressures required to initiate pouch inflation and device contraction.
Similarly, for a fixed ventricular pressure $P_v$, increasing the pouch number systematically reduces stroke volume. 

The combined effect of actuation and ventricular pressure on the fluidic lever of the device is shown in the bottom row of Figure~\ref{fig:afterload_all}.
Since stroke volume increases faster than actuation volume with a rise in actuation pressure, the device maintains the positive fluidic lever ($\Delta V_v > V_a$) for all geometries and loading considered.
We note that, for a given ventricular pressure, devices with fewer pouches exhibit higher input–output ratios due to steeper pressure–volume responses.
At constant actuation pressure, increasing ventricular pressure reduces stroke volume $\Delta V_v$, indicating afterload sensitivity.
However, increasing actuation pressure compensates for this effect, recovering stroke volume. 
This behavior suggests that simple pressure-based control strategies could maintain cardiac output across physiological afterload variations and provide a safe and reliable TAH.
We note that for $N_p=4$, and for $N_p=6$ at the highest ventricular pressure, the endocardial pouch surfaces approach a configuration that would lead to self-contact.
We indicate these configurations by reducing the line opacity. 
We do not enforce contact constraints in the present model, as they would considerably increase the computational cost and occur only in a small subset of loading conditions—mainly those with non-physiologically low afterload conditions.

We further quantify the device's mechanical efficiency by comparing the work performed by the actuation pressure in the pouches to the work delivered against the ventricular pressure.
More specifically, the efficiency of the device in converting pouch work to ventricular (stroke) work is quantified as
\begin{equation}
    \eta = \frac{W_\text{out}}{W_\text{in}} = \frac{\int P_v\,\text{d}V_v}{\int P_a\,\text{d}V_a} \,,
\end{equation}
where $P_v$ and $P_a$ are the ventricular and actuation (pouch) pressures, respectively.
Figure~\ref{fig:afterload_all} shows the efficiency for different pouch numbers $N_p$ evaluated at a physiological stroke volume of $\Delta V_v = 90$ ml in the top right panel.
Efficiencies range from 89 to 43\% for all devices and conditions tested.
Lower pouch numbers consistently yield higher efficiency, and for a given pouch number, efficiency increases with ventricular pressure (or afterload). 
\newtext{These trends are consistent with experimental observations, where the efficiency of a device with $N_p=9$ increases as ventricular pressure is increased and where devices with lower pouch numbers are significantly more efficient than devices with higher pouch numbers \cite{Arfaee2025AHeart}.
The experimentally measured efficiencies for $N_p \in \{5,7,9,11\}$ at $P_v = 3.8$ kPa are included in Figure~\ref{fig:afterload_all} for reference and show close agreement with our numerical results.
}

\newtext{Having established that our computational framework reproduces the global pressure–volume and afterload sensitivity behavior of the LIMO prototypes, we now further exploit our simulations to examine local stresses, strain paths, fatigue risk, and to perform a first exploratory assessment on simple design modifications that are costly to explore experimentally.}

%%%%%%%%%%%%%%%%%%%%%%%%%%%%%%%%
\subsection{Local stress concentration and fatigue risk}
\label{ssec:res_stress_fatigue}
\begin{figure}
    \centering
    \includegraphics[width=\textwidth]{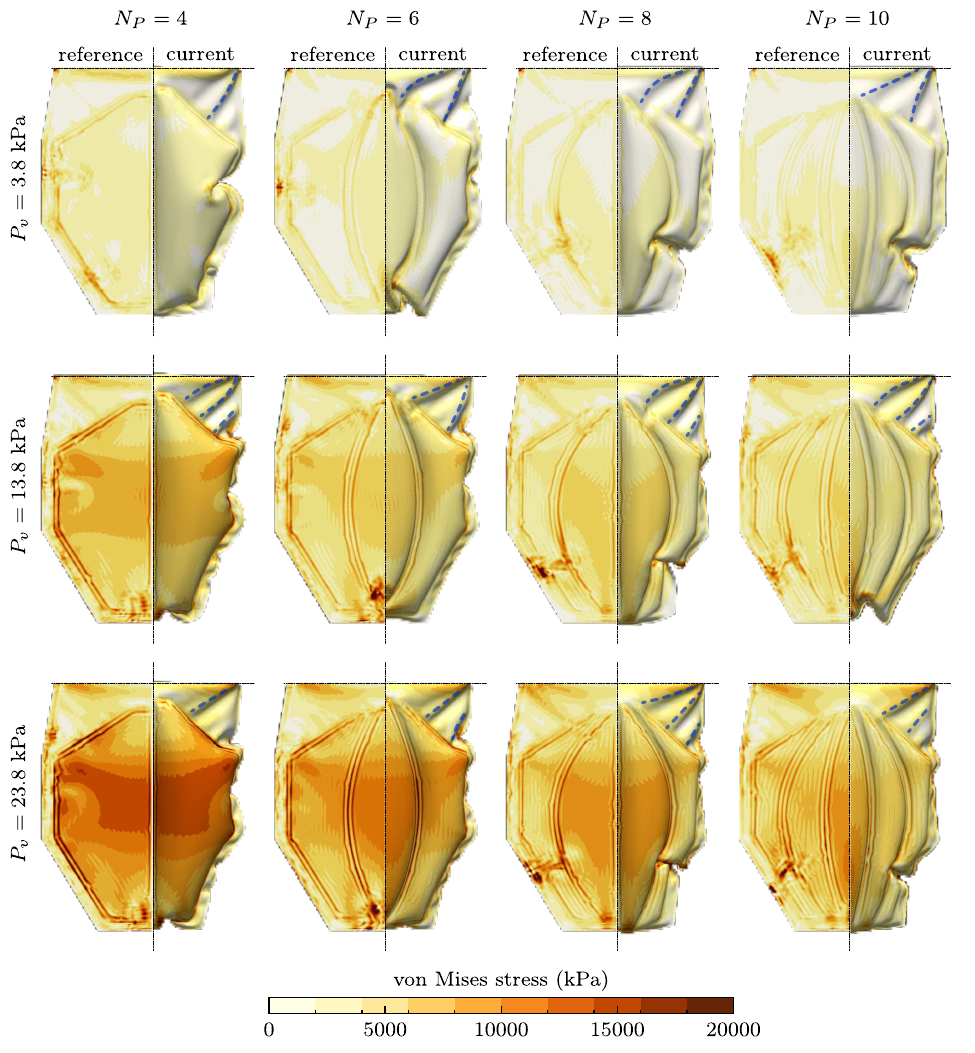}
    \caption{\textbf{Local stress concentrations across pouch numbers and ventricular pressures.} 
    Von Mises stress distribution at $P_a=P_{a,\text{max}}$ for pouch numbers $N_p \in \{4,6,8,10\}$ (left to right) and ventricular pressures $P_v \in \{3.8,13.8,23.8\}$ kPa (top to bottom).  
    To visualize the stress distribution in high detail, we project the current stress state at $P_a=P_{a,\text{max}}$ both onto the original reference geometry and the current deformed geometry, shown on the left and right side of each subplot, respectively.  
    High-stress regions align with heat-sealed seam lines and zones of pronounced buckling.  
    Persistent tension lines connecting the valve support to the upper pouch edges (dashed blue lines) are consistent with experimental observations \cite{Arfaee2025}.} 
    \label{fig:local_stress_all}
\end{figure}

%%%%%%%%%%%%%%%%%%%%%%%%%%%%%%%%
Localized buckling during inflation generates pronounced local stress concentrations, which also influence the global pressure-volume response. 
We quantify these stress concentrations using the von Mises equivalent stress, which provides a consistent scalar measure of distortional stress for the constitutive material model employed here.
Figure~\ref{fig:local_stress_all} showcases von Mises stress distributions at an actuation (pouch) pressure $P_a\in\{60,100,140\}$ kPa for each ventricular pressure ($P_v\in\{3.8,13.8,23.8\}$ kPa) and pouch number.
Across all configurations, stresses concentrate along heat-sealed seam lines and in regions of large curvature associated with localized buckling of the fabric.
\newtext{Additional simulations reported in \ref{appendix:mesh_convergence} and \ref{appendix:biffurcation} demonstrate that these dominant buckling zones and stress hotspots are robust with respect to mesh density and symmetry: even doubly symmetric meshes evolve toward similar non-axisymmetric buckling patterns and nearly identical stroke volumes.}

A second recurring feature present is the formation of distinct \emph{tension lines} between the valve support edge and upper face of the pouches.
These tension lines originate from the pre-stretched initial valve-fitting device configuration and persist throughout pouch inflation, acting as preferential load paths that connect the valve ring and pouch edges.
These tension lines are also observed experimentally \cite{Arfaee2025} (see the original publication's Figure~3b - left) and motivate our first exploratory design alteration study in Section~\ref{ssec:valve_support} where we investigate the effect of valve support geometry on local stress concentrations and device performance.

Increasing ventricular pressure $P_v$ raises the average stress across the whole device.
However, peak von Mises stresses are only weakly affected by the ventricular pressure $P_v$, as they occur primarily in high-curvature buckled regions.
For a fixed ventricular pressure, devices with fewer pouches exhibit higher peak stresses, reflecting the larger load carried per seam and greater buckling amplitudes, whereas devices with more pouches distribute the load over a higher number of seams and exhibit reduced buckling severity.

\begin{figure}
    \centering
    % \resizebox{\textwidth}{!}{\input{figs/limo_fatigue.pgf}}
    \includegraphics[width=\textwidth]{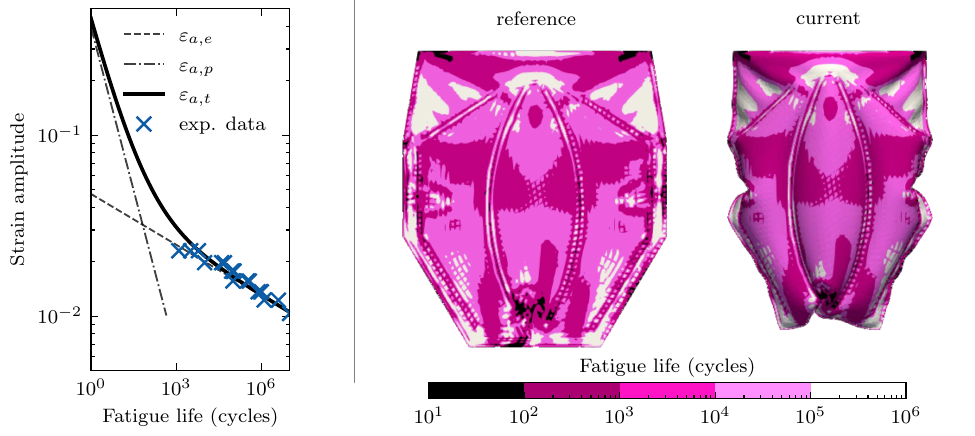}
    \caption{\textbf{Fatigue life risk index for the device.} (left) Strain-life curve for a pure TPU polymer with plastic and elastic parts and the experimental data used to calibrate the elastic part of the strain-cycle curve.
    (right) Isocontours of the expected number of cycles of a \textbf{pure} TPU device using the maximum principal strain in the membrane as a reference strain for the fatigue assessment.
    Note that we assume that the weakest points are near the seams on the TPU-TPU bonds, and thus we use a strain-cycle curve for pure TPU only.
    The expected number of cycles is only valid in the seams region.}
    \label{fig:fatigue_life}
\end{figure}

To assess fatigue risk, we adopt a simplified strain-life (E–N) approach.
\newtext{Based on experimental observations that LIMO device failure consistently initiates at TPU-TPU heat-sealed seams rather than within the combined TPU-nylon fabric itself \cite{Arfaee2025}, we evaluate fatigue using strain–life data for bulk TPU as a first-order proxy for seam behavior. While this choice does not capture the detailed mechanics of bonded seams, it provides a consistent basis for comparative fatigue-risk mapping, and predicted cycle counts should therefore be interpreted as indicative rather than predictive.}

Complete strain–life curves are commonly modeled with the \emph{Coffin–Manson–Basquin} relation, which links the total strain amplitude $\varepsilon_{a,t}$ to the number of loading cycles $N_\text{f}$ to failure:
\begin{equation}
    \varepsilon_\text{a,t} = \varepsilon_\text{a,e} + \varepsilon_\text{a,p} = \frac{\sigma_\text{f}'}{E}(2N_\text{f})^b + \varepsilon_\text{f}'(2N_\text{f})^c.
\end{equation}
Here, $\varepsilon_\text{a,e}$ and $\varepsilon_\text{a,p}$ denote the elastic and plastic strain amplitudes, respectively.
The Young’s modulus $E$ of pure TPU material is typically obtained from a static tensile test or the first hysteresis loop of a fatigue test \cite{Niesony2008NewCompatibility}, and 
$\sigma_\text{f}'$, $\varepsilon_\text{f}'$, $b$, and $c$ represent the fatigue strength coefficient, fatigue ductility coefficient, fatigue strength exponent, and fatigue ductility exponent, respectively.
Direct experimental determination of these parameters for the specific seam material used in the LIMO device is beyond the scope of this work. 
Instead, as shown in Figure~\ref{fig:fatigue_life} - left, we inform the elastic part of the curve $\varepsilon_{\text{a},e}$ based on reported TPU-parameters \cite{Wang2023CyclicFatigue}.
We construct a compatible plastic branch by enforcing $\varepsilon_{\text{a},p}<\varepsilon_{\text{a},e}$ for $N_\text{f}<10^2$ and using a conservative fatigue ductility coefficient $\varepsilon_\text{f}'=0.55$ based on the yield strain of TPU \cite{Ayedi2025EnhancingEffects}.
Given that the device operates predominantly in the elastic regime under the considered loading conditions, this calibration is sufficient for a first-order fatigue screening.
As a conservative measure, we use the maximum principal strain attained during inflation as the strain amplitude input to the E-N relation at each material point.
The resulting estimates of $N_\text{f}$ define a fatigue risk field across the whole device geometry.
Figure~\ref{fig:fatigue_life} - right shows the predicted fatigue life index. Corresponding cycles-to-failure proxy estimates across the other geometries and ventricular pressures are provided in ~\ref{appendix:fatigue_all}.
The shortest predicted lives are found along the heat-sealed seams and along the persistent tension lines identified in Section~\ref{ssec:res_stress_fatigue}, coinciding with regions of elevated von Mises stress. \newtext{In particular, the coincidence of high von Mises stress, short predicted fatigue life, and experimentally observed seam failures indicates that the present framework can rationalize past failures and guide where geometric modifications or local reinforcements are likely to be most effective.} 

%%%%%%%%%%%%%%%%%%%%%%%%%%%%%%%%
\subsection{Principal strain fields}
\label{sec:strain_path}
%%%%%%%%%%%%%%%%%%%%%%%%%%%%%%%%
Since the LIMO device is constructed from a woven fabric whose fiber orientation could be tailored in future prototypes \cite{Nagy2013IsogeometricDistribution}, we analyze the principal strain fields to identify dominant load paths and potential directions for anisotropic reinforcement \cite{Belforte2014BellowsMuscle, Connolly2015MechanicalAngle}.
We obtain principal strains by computing the eigenvalues $E_i$ of the Green–Lagrange strain tensor $\ten{E}$:
\begin{equation}
    \ten{E}\vec{v}_i = E_i \vec{v}_i\,, \quad \to \quad (\ten{E} - E_i \mathbb{I}) \vec{v}_i = \vec{0} \qquad (i=1,2,3;\,\, \text{no summation}),
\end{equation}
with $E_1$, $E_2$, and $E_3$ corresponding to the maximum, intermediate, and minimum principal strains, respectively, and $\vec{v}_i$ is the associated eigenvector (principal direction).

\begin{figure}
    \centering
    \includegraphics[width=\textwidth]{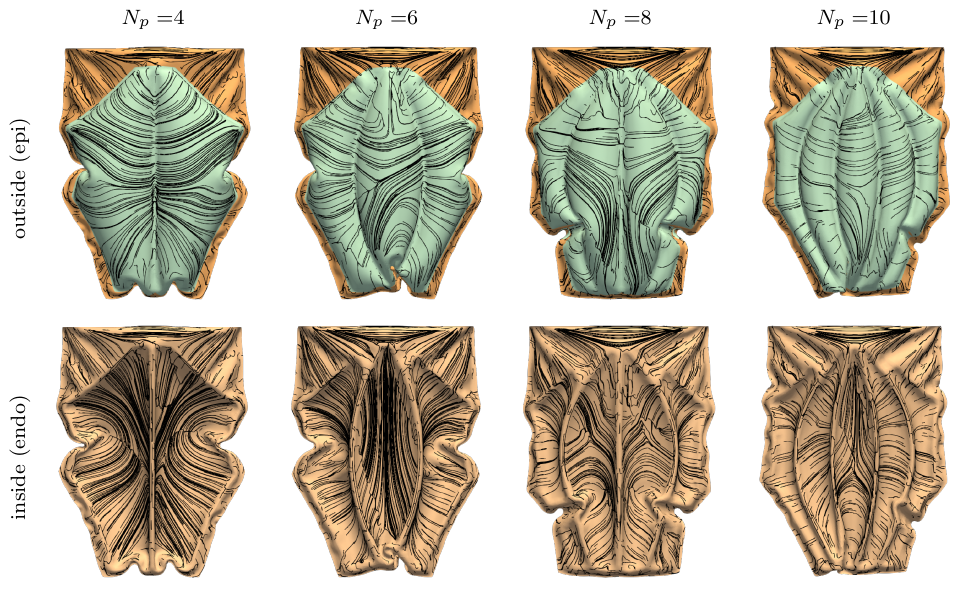}
    \caption{\textbf{Inner and outer principal strain directionality in the soft artificial heart design.} Direction of the maximum principal strain $\vec{v}_{1}$ represented as surface streamlines on the device for different numbers of pouches $N_p$ under a  ventricular pressure $P_v=13.8$ kPa. The different columns are for different pouch numbers, the top row shows a view from the outside surface (epicardium), and the bottom row shows a view of the inner surface (endocardium).}
    \label{fig:strain_map}
\end{figure}

Figure~\ref{fig:strain_map} visualizes the maximum principal strain direction $\vec{v}_{1}$ on the outer epicardial and inner endocardial surfaces of the device. 
Given the isotropic material model, these principal strain directions naturally coincide with the principal stress directions.
On the outer epicardial pouch surface, the maximum principal strain is predominantly circumferential, consistent with resisting global ballooning of the pouches and device.
In contrast, on the endocardial pouch surface, the maximum principal strain is oriented approximately orthogonal to the epicardial direction, indicating that the inner fabric mainly pulls inward and upward upon pouch inflation.
At the seam lines, where the two pouch fabric surfaces merge, the principal strain field aligns across the thickness. 
These locations coincide with peak von Mises stresses and the persistent tension lines identified in Section~\ref{ssec:res_stress_fatigue}, further supporting the role of the seams as critical load-bearing paths and potential fatigue initiation sites.

%%%%%%%%%%%%%%%%%%%%%%%%%%%%%%%%
%%%%%%%%%%%%%%%%%%%%%%%%%%%%%%%%
\section{In silico design exploration}
%%%%%%%%%%%%%%%%%%%%%%%%%%%%%%%%
%%%%%%%%%%%%%%%%%%%%%%%%%%%%%%%%
\subsection{Effect of valve support geometry}
\label{ssec:valve_support}
\begin{figure}[!ht]
    \centering
    % \resizebox{\textwidth}{!}{\input{figs/limo_pressure_vonmises_AR.pgf}}
    \includegraphics[width=\textwidth]{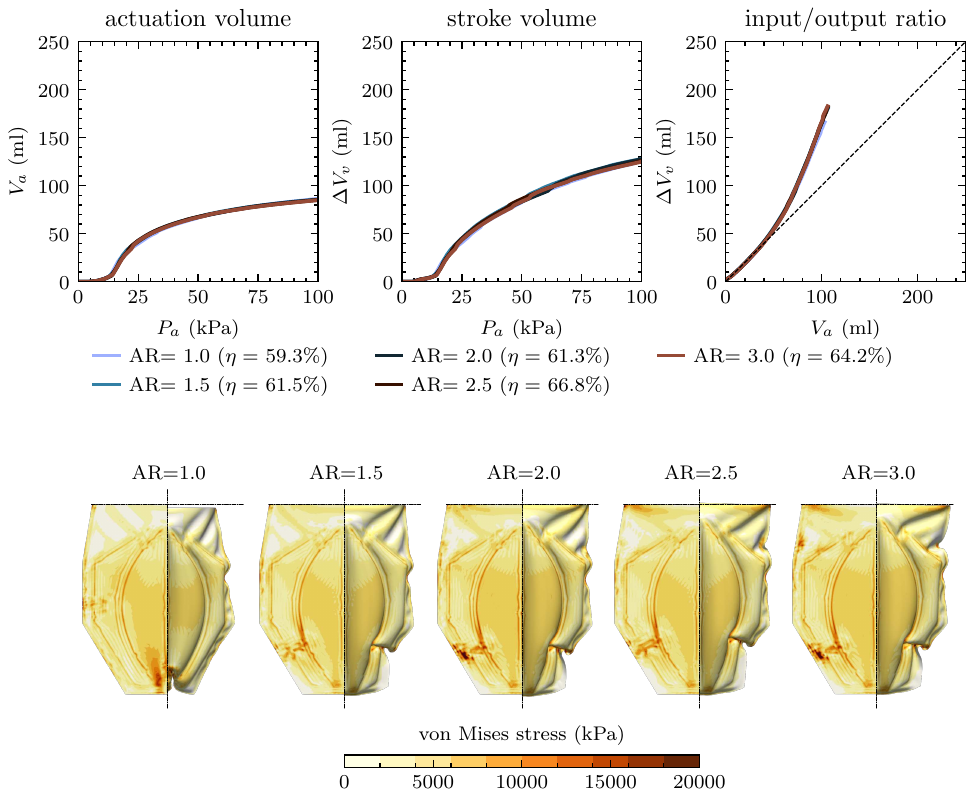}
    \caption{\textbf{Effect of valve support geometry on global performance and local stress concentrations.}
    Pressure–volume response and von Mises stress distributions for varying valve support aspect ratios under a ventricular pressure of $P_v = 13.8$ kPa.
    Top row: actuation (pouch) volume versus actuation pressure (left), stroke volume versus actuation pressure (center), and input–output ratio (right); mechanical efficiencies are indicated in the legend.
    Bottom row: von Mises stress distributions at $P_a = P_{a,\text{max}}$ for different valve support aspect ratios (left to right), shown on both the reference and deformed configurations.
    }
    \label{fig:results_AR}
\end{figure}

\newtext{
Following our computational and experimental observations of persistent \textit{tension lines} in the original LIMO designs \cite{Arfaee2025}, we use our computational framework to examine how variations in the geometry of the rigid valve support influence local stress concentrations and global device performance.
Specifically, we vary the valve support's aspect ratio $\text{AR}\in\{1.0,1.5,2.0,2.5,3.0\}$ and consider a representative ventricular pressure $P_v=13.8$ kPa, corresponding to physiological diastolic pressures in the systemic outflow tract.
As shown in Figure~\ref{fig:results_AR} - top, variations in valve support geometry have little influence on stroke volume during pouch inflation. 
Across all aspect ratios, we find that the LIMO design's intrinsic fluid lever effect is preserved (Figure~\ref{fig:results_AR} - top right), and the mechanical efficiency remains within a modest range, $\eta \in \{59.3\%,66.8\%\}$, with a modest peak at $AR=2.5$.
}

\newtext{
In contrast, the valve support aspect ratio has a pronounced effect on local stress concentrations.
Figure~\ref{fig:results_AR} - bottom shows the resulting von Mises stress map in the reference and current configuration.
While global performance metrics such as the efficiency and \textit{Less In More Out} effect are largely unaffected,  by the valve support's aspect ratio, reducing the aspect ratio toward a more circular support leads to a clear reduction in peak stress.
Compared to the most slender design ($\text{AR}=3.0$), smaller aspect ratios yield an 8\% reduction in the peak von Mises stress, quantified as the area-averaged top 5\% of stress values.
Additionally, lower valve support aspect ratios also reduce stress levels near the attachment of the device onto the support.
Taken together, these results show that a modest change in valve support aspect ratio can reduce peak von Mises stress by nearly 10\% while preserving the fluidic lever and stroke volume, demonstrating how the model can identify low-cost geometric modifications with a potentially large impact on durability.}

\subsection{Effect of relative endocardial-epicardial compliance}
\newtext{
The physical device and baseline numerical model assume identical mechanical properties for the endocardial and epicardial pouch fabrics. 
However, Section~\ref{sec:strain_path} showed that these two surfaces experience distinct strain paths, which suggests that differentiating their mechanical properties may be exploited to optimize device performance.
Here, we investigate the effect of reducing the stiffness of the endocardial pouch surface relative to the epicardium and the rest of the device.}

\begin{figure}
    \centering
    % \resizebox{\textwidth}{!}{\input{figs/limo_pressure_vonmises_Ehat.pgf}}
    \includegraphics[width=\textwidth]{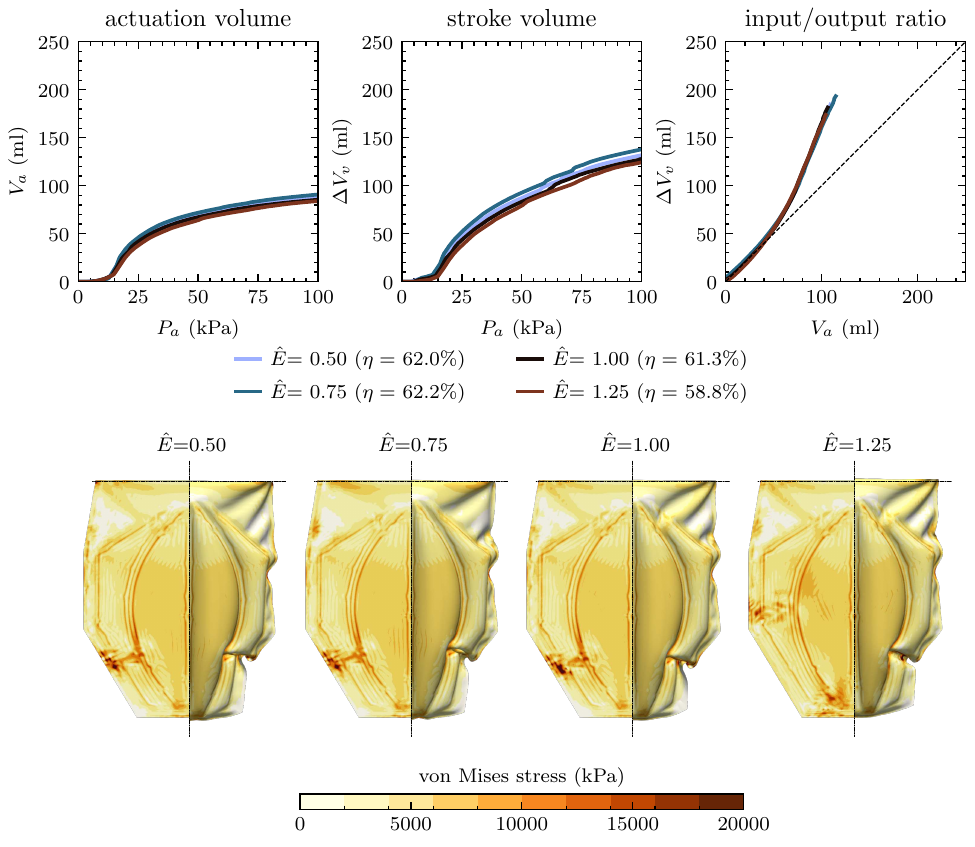}
    \caption{\textbf{Effect of relative endocardial–epicardial compliance on global performance and local stress concentrations.}
    Pressure–volume response and von Mises stress distributions for varying relative endocardial stiffness scaling factors $\hat{E}$ under a ventricular pressure of $P_v = 13.8$ kPa.
    Top row: actuation (pouch) volume versus actuation pressure (left), stroke volume versus actuation pressure (center), and input–output ratio (right); mechanical efficiencies are indicated in the legend.
    Bottom row: von Mises stress distributions at $P_a = P_{a,\text{max}}$ for decreasing endocardial stiffness (left to right), shown on both the reference and deformed configurations.
    }
    \label{fig:results_Ehat}
\end{figure}

\newtext{We vary the ratio of endocardial pouch stiffness to the reference device stiffness,
\begin{equation}
    \hat{E}\equiv E_\text{endo}/E \in \{0.50,0.75,1.00,1.25\},
\end{equation}
and evaluate device performance under a ventricular pressure of $P_v = 13.8$ kPa.
Figure~\ref{fig:results_Ehat} - top shows the resulting stroke volume for different stiffness ratios $\hat{E}$.
Lower stiffness ratios lead to increased actuation and stroke volumes, as less elastic energy is required to deform the pouches.  
The highest stroke volume is obtained for $\hat{E} = 0.75$ (Figure~\ref{fig:results_Ehat} - top center).
}

\newtext{
The fluidic lever is only weakly affected by changes in endocardial stiffness; however, mechanical efficiency increases as the stiffness ratio is reduced, with a maximum again observed at $\hat{E} = 0.75$. Because the local stiffness distribution is altered, peak von Mises stresses also change relative to the reference design ($\hat{E} = 1.00$), as shown in Figure~\ref{fig:results_Ehat} - bottom. Lower stiffness ratios reduce peak stresses by approximately 10\%, whereas increasing the stiffness ratio leads to peak stress levels up to 8\% higher than the reference configuration and produces different deformation patterns.
This simple material-contrast study illustrates that tuning the relative endocardial–epicardial compliance provides an additional design degree of freedom: selectively softening the inner pouch surface increases stroke volume and reduces peak stresses without altering overall geometry or actuation strategy.
}

% \clearpage
%%%%%%%%%%%%%%%%%%%%%%%%%%%%%%%%
\section{Discussion}
\label{sec:discussion}
%%%%%%%%%%%%%%%%%%%%%%%%%%%%%%%%

\textbf{Our framework reproduces the nonlinear pressure--volume behavior and inflation onset observed experimentally.}
We developed a numerical framework to investigate the mechanics of the \textit{Less In, More Out} (LIMO) soft robotic artificial heart device and \newtext{replicated the static inflation experiments reported in Arfaee\textit{~et~al.}~(2025)\cite{Arfaee2025} 
}
Across all tested geometries and ventricular pressures, the model captures the device's characteristic nonlinear pressure--volume response, including the onset of pouch inflation once the actuation pressure exceeds the ventricular pressure and the subsequent reconfiguration-driven increase in stroke volume (Fig.~\ref{fig:pressure_vs_volume}).
Devices with fewer pouches generate larger stroke volumes at identical actuation pressures, while all configurations achieve an output--input ratio greater than one, consistent with experimental observations.
\newtext{Quantitative discrepancies between numerical and experimental results emerge primarily for higher pouch numbers, where bending-dominated deformation becomes more prominent.
In this regime, the numerical pressure--volume response deviates modestly from experiments during the reconfiguration phase, while agreement improves again at higher actuation pressures where volume changes become small (Figs.~\ref{fig:deformation_aft0} and \ref{fig:pressure_vs_volume}).
These differences are primarily attributed to two modeling assumptions.
First, the use of an isotropic St.\ Venant--Kirchhoff material law, calibrated solely to uniaxial tensile data, cannot capture the anisotropic in-plane response of woven fabrics.
Second, the adoption of a continuum shell formulation links the bending and the tensile modulus to the Young's (elastic) modulus measured via in-plane tensile tests. For our TPU-coated woven textile, this assumption overestimates the bending compliance of the material \cite{Teng1999}.
Indeed, woven textiles exhibit highly nonclassical bending behavior governed by yarn rotation, inter-yarn slippage, frictional locking, and local micro-buckling, phenomena that are well documented across filament-, yarn-, and fabric-scale studies \cite{King2005AFabrics, Erol2017ACell, Hu2004StructureFabrics}.
These mechanisms control both the in-plane response and the apparent bending compliance of the fabric through progressive yarn reorientation and micro-buckling.
In-plane uniaxial tensile testing data alone cannot calibrate a woven fabric material model\cite{Belforte2014BellowsMuscle, Arfaee2023ModelingMotors, Connolly2019Sew-freeRobots, Guo2024EncodedRobots}, and might require separate bending modulus calibration \cite{Teng1999}.
In the absence of comprehensive multi-axial experimental data, this limitation has nevertheless motivated the widespread adoption of effective isotropic continuum models as first-order approximations in fabric-based soft robotic actuators, including the present study.
As a result, deformation modes involving significant out-of-plane bending, which is most prominently observed in devices with higher pouch numbers, are predicted with reduced quantitative accuracy in the present framework.
Addressing these modeling limitations requires two complementary advances.
Fabric-specific multi-axial characterization and constitutive descriptions are needed to improve the in-plane response, for which coupling targeted experimental data with emerging data-driven and physics-informed material discovery approaches \cite{Peirlinck2024a, McCulloch2024AutomatedFabrics, Thakolkaran2025CanCANs} offers a promising route.
In parallel, enriched TPU-coated fabric material bending stiffness descriptions and experiments will be required to accurately represent the bending-dominated behavior of woven fabrics.
Geometric front--back symmetry and other idealizations, including the omission of discontinuous pouch connections, actuation inlets, and the peripheral edge skirt, are expected to have a secondary influence on the global pressure--volume response.
However, these features do introduce local stiffness heterogeneities and stress concentrations that may affect detailed deformation patterns and fatigue life.
}

\textbf{Our models capture afterload sensitivity and efficiency trade-offs across pouch numbers.}
Soft artificial hearts must operate robustly across a wide range of hemodynamic loading conditions.
Our simulations reproduce the experimentally observed afterload sensitivity of the LIMO device: increasing ventricular pressure shifts the onset of inflation to higher actuation pressures and reduces stroke volume at fixed actuation pressure, while increasing actuation pressure compensates for this effect (Fig.~\ref{fig:afterload_all}).
Mechanical efficiency increases with ventricular pressure, in agreement with experimental measurements \cite{Arfaee2025}, and remains strongly dependent on pouch number.
Nevertheless, the present efficiency estimates are obtained under quasi-static loading and neglect fluid inertia.
In practice, inertia effects may dampen rapid buckling events and slightly modify instantaneous work exchange during the cardiac cycle.
\newtext{Future extensions of the framework will therefore couple the structural model to lumped-parameter circulation models \cite{Peirlinck2018b, Peirlinck2021,Regazzoni2022ACirculation} and fully three-dimensional fluid solvers, for example using immersed-boundary formulations \cite{Lauber2022, Bornoff2023, Lauber2024Immersed-BoundaryShells}
Such coupled simulations will enable resolution of unsteady flow features, including recirculation, stagnation, and shear exposure \cite{Park2022ComputationalFunction, Singh2023RoboticIntervention, Roche2025IntegratingFibrillation, Bakker2025}, and support optimization under physiologically realistic cardiac cycles.}

\textbf{Full-field stresses and strain directions reveal durability-limiting load paths.}
Durability remains a central challenge for fabric-based soft robotic devices.
By resolving full-field stresses, the present model identifies pronounced von Mises stress concentrations along heat-sealed seams and in regions of localized buckling, consistent with experimentally observed failure locations (Fig.~\ref{fig:local_stress_all}).
Beyond stress magnitude, the principal strain directions reveal a distinct and systematic pattern: outer pouch surfaces deform predominantly circumferentially to resist ballooning, while inner surfaces experience axial stretching approximately orthogonal to the epicardial strain direction (Fig.~\ref{fig:strain_map}).
This kinematic organization mirrors key aspects of native left ventricular mechanics, where opposing fiber orientations across the wall thickness contribute to torsion and efficient ejection \cite{Sengupta2008TwistApplication, Peirlinck2022, Osouli2025}.
These findings suggest that tailoring fabric anisotropy, fiber orientation, or local reinforcement along dominant strain paths offers a promising strategy to further enhance performance and durability \cite{Bui2023}
While the present study employs an isotropic effective material model appropriate for TPU-coated fabrics, future designs may explicitly exploit woven anisotropy to more closely emulate physiological deformation modes.

\textbf{Our strain--life fatigue screening identifies seam- and buckling-driven failure risks.}
\newtext{We implemented a simplified strain--life fatigue assessment based on a Coffin--Manson--Basquin formulation calibrated to bulk TPU data from the literature \cite{Wang2023CyclicFatigue}.
Although this approach does not yield absolute lifetime predictions for bonded fabric seams, it provides a consistent first-order framework for identifying relative fatigue risk across device geometries.
Regions predicted to experience the shortest fatigue lives coincide with seam lines and persistent tension paths (Figs.~\ref{fig:local_stress_all} and \ref{fig:fatigue_life}), aligning with experimental observations of failure initiation \cite{Arfaee2025}.}
Importantly, the coincidence of high von Mises stress, elevated principal strain amplitudes, and experimentally observed failure locations indicates that the framework captures the dominant mechanical drivers of fatigue.
\newtext{As targeted strain--life data for bonded TPU-fabric seams become available, the present workflow can be readily refined toward quantitatively predictive lifetime estimates.}

\textbf{Targeted geometric and material modifications reduce peak stress without sacrificing performance.}
By integrating global performance metrics with local stress and strain information, the framework provides a unified view of intrinsic device mechanics.
\newtext{As a proof of concept, we investigated geometric and material design modifications that are difficult to explore experimentally.
Reducing the valve support aspect ratio lowers peak von Mises stress by nearly 10\% while preserving stroke volume and mechanical efficiency (Fig.~\ref{fig:results_AR}).
Similarly, selectively softening the endocardial pouch surface increases stroke volume and reduces peak stress without altering overall geometry or actuation strategy (Fig.~\ref{fig:results_Ehat}).
Overall, these results demonstrate how simulation-driven insight can identify low-cost design modifications with important benefits for durability.
Beyond the LIMO device, the framework is directly applicable to a broad class of fabric-based, fluidically actuated soft robotic systems.
Coupled with physiologically realistic loading and patient-specific boundary conditions, it offers a systematic pathway toward the rational design of next-generation soft robotic artificial hearts that balance efficiency, durability, and hemodynamic safety.}

%%%%%%%%%%%%%%%%%%%%%%%%%%%%%%%%
\section{Conclusion}
%%%%%%%%%%%%%%%%%%%%%%%%%%%%%%%%
We have developed a computational framework for the mechanical analysis of a fabric-based, fluidic actuated soft robotic, total artificial heart and benchmarked it against published static inflation experiments. 
Our model reproduces key global observations, including inflation onset, nonlinear pressure-volume behavior, and afterload sensitivity, while also resolving full-field principal strain field directionality and local stress concentrations that are difficult to measure experimentally. 
Using a first-order strain–life screening approach, we further provide relative fatigue-risk maps that identify seams, tension-line load paths, and buckling zones as durability-limiting features.
\newtext{As proof-of-concept design studies, we show that (i) reducing the valve-support aspect ratio decreases peak von Mises stress by nearly 10\% without reducing stroke volume, and (ii) selectively softening the endocardial pouch surface increases stroke volume while lowering peak stress by $\sim10\%$.}
While demonstrated for a soft total artificial heart, our workflow is transferable to a broad class of fabric-based soft robotic and provides a physics-based route toward geometry and material tailoring, targeted reinforcement placement, and improved durability under application-specific loading conditions.

%%%%%%%%%%%%%%%%%%%%%%%%%%%%%%%
\section*{Acknowledgements}
%%%%%%%%%%%%%%%%%%%%%%%%%%%%%%%%
The authors gratefully acknowledge the support of J.T.B. Overvelde (AMOLF, The Netherlands).
This publication is part of the project Holland Hybrid Heart with file number NWA.1518.22.049 of the research program Onderzoek op Routes door Consortia 2022 – NWA-ORC 2022, which is financed by the Dutch Research Council (NWO), the Dutch Ministry of Education, Culture and Science (OCW), and the Hartstichting (Dutch Heart Foundation). 

\clearpage
\appendix
\setcounter{figure}{0} % reset figure counter
%%%%%%%%%%%%%%%%%%%%%%%%%%%%%%%%%%%%%%%%%%%%%%%%%%%%%%%%%%%%%%%%
%%%%%%%%%%%%%%%%%%%%%%%%%%%%%%%%%%%%%%%%%%%%%%%%%%%%%%%%%%%%%%%%
%%%%%%%%%%%%%%%%%%%%%%%%%%%%%%%%%%%%%%%%%%%%%%%%%%%%%%%%%%%%%%%%

\section{Material model  calibration}
\label{appendix:calibration}

\begin{figure}[!hb]
    \centering
    \includegraphics[width=0.4\textwidth]{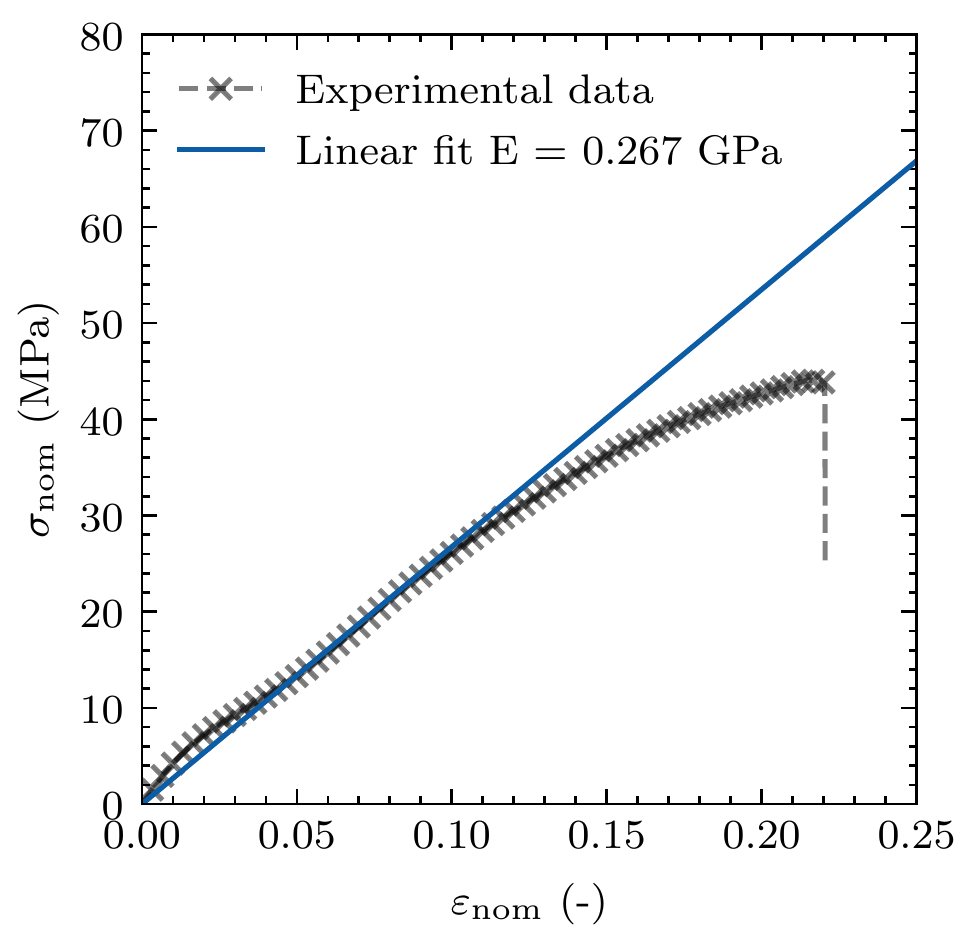}
    \caption{\textbf{Material model fitting for the nylon-coated TPU material.}
     Uniaxial nominal stress–strain relationship for a sample of the fabric used in the LIMO device, together with the linear elastic fit used to calibrate the Young’s modulus. Only strains $\varepsilon_\text{nom}<0.1$ are used in the fit, corresponding to the expected operative strain range in the device.}
    \label{fig:material_model}
\end{figure}

We calibrate our St.\ Venant--Kirchhoff material model based on experimental tensile tests of the nylon-coated TPU fabric used in the LIMO device. 
The data are obtained from uniaxial extension tests on a coupon of initial length $l_0 = 200~\text{mm}$, width $w_0 = 50~\text{mm}$, and thickness $t_0=0.19~\text{mm}$. 
From the measured force $F$ and initial cross-sectional area $A_0$ we compute the nominal (engineering) stress 
$\sigma_\text{nom} = F/A_0$, and from the measured elongation we compute the engineering strain 
$\varepsilon_\text{nom} = (l - l_0)/l_0$, with the corresponding stretch $\lambda = 1 + \varepsilon_\text{nom}$.
Over the operative strain range of interest $\varepsilon_\text{nom} < 0.1$, the Green--Lagrange strain in the loading direction,
\[
    E_{11} = \tfrac{1}{2}(\lambda^2 - 1)
           = \varepsilon_\text{nom} + \tfrac{1}{2}\varepsilon_\text{nom}^2,
\]
differs from $\varepsilon_\text{nom}$ only by a quadratic correction; thus $\varepsilon_\text{nom}$ provides a good approximation to $E_{11}$ in this regime.
We therefore identify the slope of the nominal stress--strain curve with the effective Young’s modulus $E$ from a least-squares fit
\begin{equation}
    \min_{E} \sum_{i=1}^{N} \left\lvert \sigma_{\text{nom},i} - E\,\varepsilon_{\text{nom},i} \right\rvert^{2},
\end{equation}
which yields $E = 0.267~\text{GPa}$. 
Combined with an assumed Poisson’s ratio $\nu = 0.33$, this Young’s modulus defines the Lamé parameters used in our St.\ Venant--Kirchhoff constitutive model (Eq.~\ref{eq:stvenantkirchoff}).
The experimental data and the fitted linear curve over the calibration range are shown in Figure~\ref{fig:material_model}.

\clearpage
%%%%%%%%%%%%%%%%%%%%%%%%%%%%%%%%
\section{Mesh sensitivity study}
\setcounter{figure}{0} % reset figure counter
%%%%%%%%%%%%%%%%%%%%%%%%%%%%%%%%

%%%%%%%%%%%%%%%%%%%%%%%%%%%%%%%%
\subsection{Square airbag inflation convergence}
\label{appendix:solver_validation}
%%%%%%%%%%%%%%%%%%%%%%%%%%%%%%%%
We validate our finite element implementation by simulation the inflation of a square airbag.
This test case has been widely used in previous work to validate numerical models of inflated membrane-like structures using either finite element methods \cite{Cirak2001FullyAnalysis, Ziegler2003AStructures, Lee2006FiniteDeformations} or isogeometric analysis \cite{Chen2014, Nakashino2020}.

\begin{figure}[!ht]
    \centering
    \includegraphics[width=\textwidth]{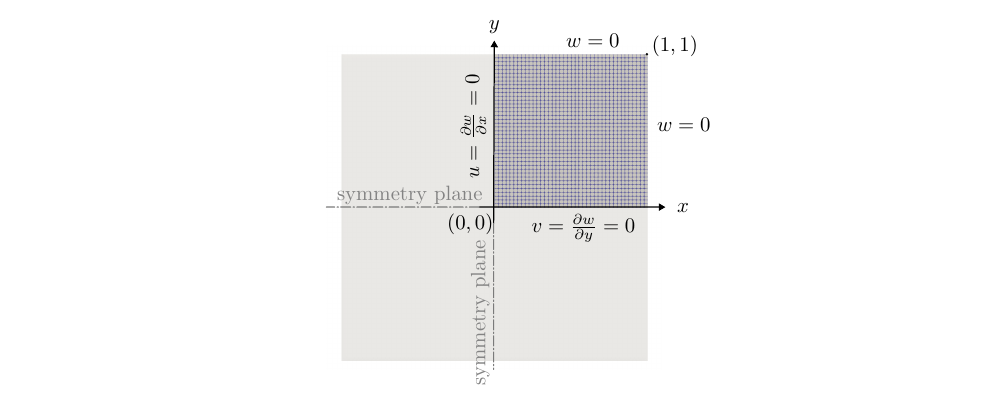}
    \caption{\textbf{Square airbag validation study: schematic of the used domain, boundary conditions and mesh.}
    Symmetric boundary conditions are imposed in the symmetry planes ($x=0$ and $y=0$) while the vertical displacement of the outer-edge is fixed ($w=0$).
    Since we have three symmetry planes, we model only 1/8$^\text{th}$ of the geometry, here represented by a quadrilateral mesh with $N_e=41$, instead of the total upper surface of the airbag (grey area).}
    \label{fig:square_airbag_geom}
\end{figure}

This problem setting involves an initially flat square airbag with a side length of $\sqrt{1/2}$ and a thickness of 0.1 mm, which is inflated by a uniform internal pressure load.
The Young’s modulus and Poisson’s ratio of the membrane are $E=2$ GPa and $\nu=0.3$, respectively, and we prescribe a uniform pressure load of 5.00 kPa.
For this problem, we leverage symmetry to simplify the problem to modeling only one eighth of the membrane (i.e., one quarter of the upper half of the square airbag).
The boundary conditions of our partial model are shown in Figure~\ref{fig:square_airbag_geom}.

\begin{figure}[!ht]
    \centering
    \includegraphics[width=\textwidth]{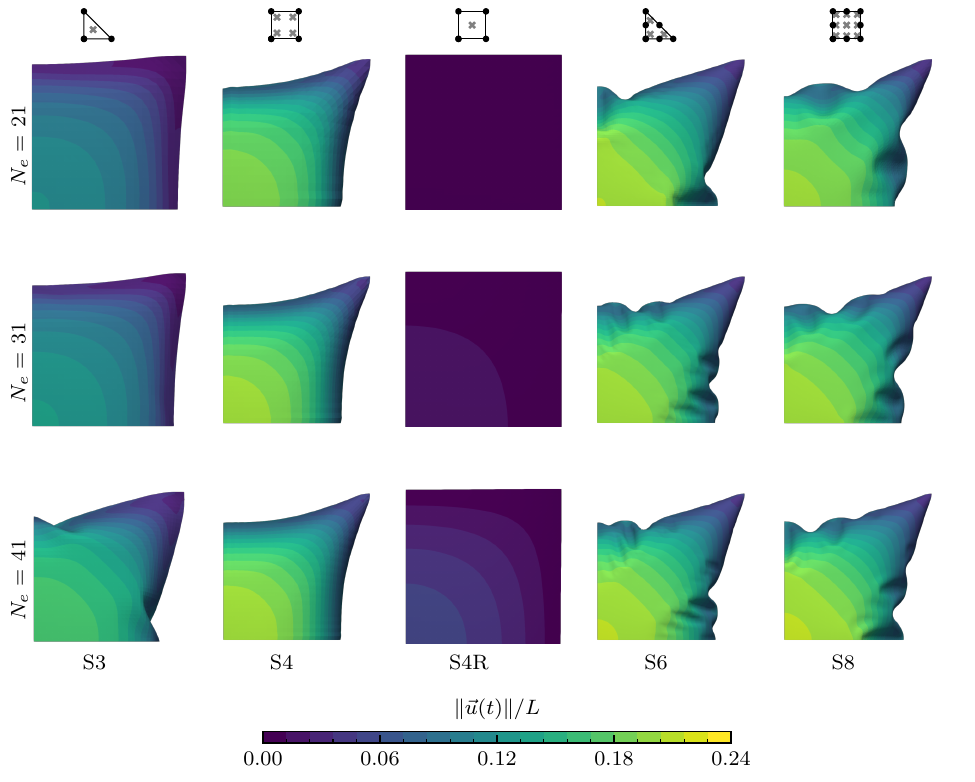}
    \caption{\textbf{Deformation of the square airbag for different mesh densities and element types.}
    Different rows represent different mesh densities (with the left column using a slightly different element count due to the poor performance of S3 elements, i.e., locking).
    Different columns represent different element types, where R stands for reduced integration.
    The displacement magnitude ($\Vert\Vec{u}\Vert/L$) iso-contours consist of 21 equally-spaced levels between $[0, 0.24]$.
    The reference element and the integration points location are depicted as small insets in the top row of the figure.}
    \label{fig:square_deformation}
\end{figure}

Figure~\ref{fig:square_deformation} shows the final deformation of the square membrane for different element types and mesh densities. 
Linear elements (S3, S4, and S4R) either lock (S3) or do not resolve the complex deformation patterns. 
Quadratic elements perform markedly better: they capture the characteristic buckling and folding modes of inflatable airbags \cite{Nakashino2020} and converge rapidly with mesh refinement. 

\begin{figure}[!ht]
    \centering
    \includegraphics[width=\textwidth]{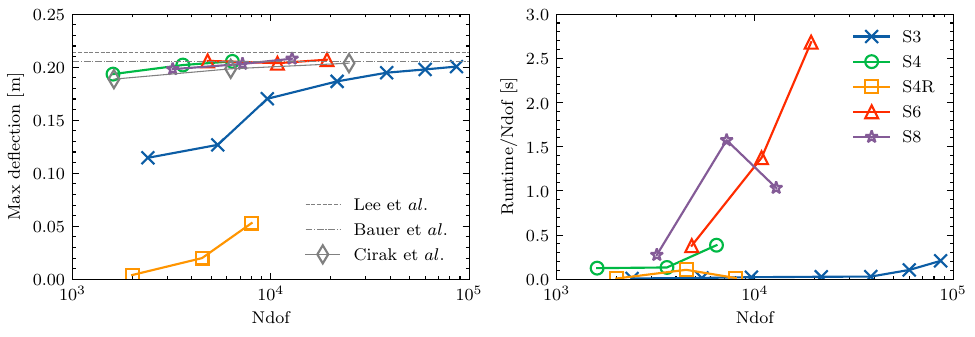}
    \caption{\textbf{Convergence and runtime for different element types and number of degrees of freedom for the square airbag inflation case.} Convergence (left) of the maximum vertical displacement and relative runtime (right) for different mesh densities and elements for the square airbag inflation case.
    Higher-order elements exhibit the expected convergence behavior, whereas S3 and S4R suffer from locking and fail to converge to the reference solutions shown in gray.}
    \label{fig:square_convergence}
\end{figure}

Figure~\ref{fig:square_convergence} quantifies mesh convergence and computational cost for the square airbag benchmark.
We track the maximum vertical displacement as a function of the number of degrees of freedom and compare it to reference values reported in the literature \cite{Bauer1975, Lee2006FiniteDeformations, Cirak2001FullyAnalysis}.
Clearly, from all the elements tested, only S4, S6, and S8 elements perform satisfactorily and are able to match the reference results.
In terms of computational cost, quadratic elements require more computational effort; however, their accuracy is also significantly higher, making them the preferred choice for this type of structure.

\clearpage
%%%%%%%%%%%%%%%%%%%%%%%%%%%%%%%%
\subsection{LIMO inflation convergence}
\label{appendix:mesh_convergence}
\begin{figure}[!ht]
    \centering
    \includegraphics[width=\textwidth]{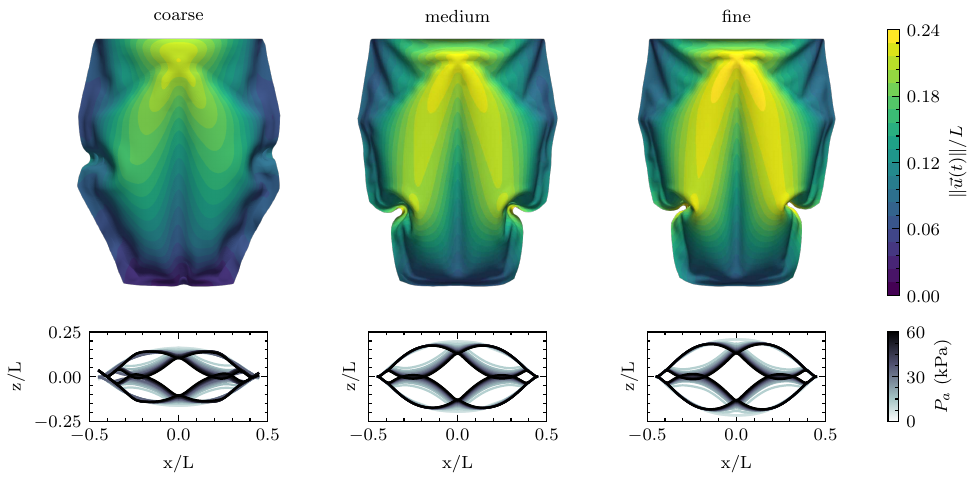}
    \caption{\textbf{Convergence study of the LIMO device deformation for $N_p=4$ under a ventricular pressure $P_v=13.8$ kPa.} 
    The top row shows the deformed geometry at peak actuation pressure, colored by deformation magnitude normalized by the reference device length, for the coarse (left), medium (center), and fine (right) meshes.
    The bottom row shows cross-sectional deformation profiles at $y/L=0.5$ extracted at 20 evenly spaced actuation pressure levels $P_a$, from the onset of pouch inflation (white) to the maximum actuation pressure (black).}
    \label{fig:limo_valid_deformation}
\end{figure}

\begin{figure}[!ht]
    \centering
    \includegraphics[width=\textwidth]{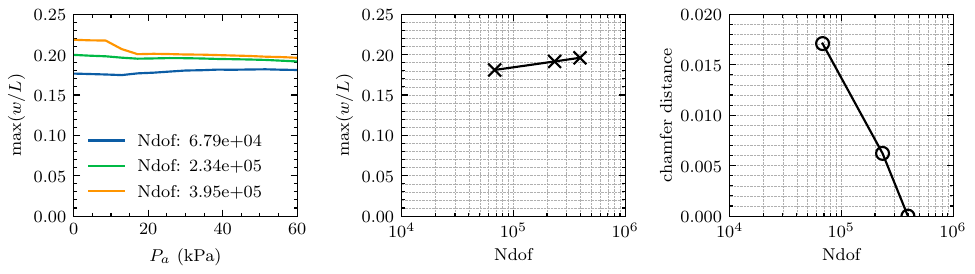}
    \caption{\textbf{Comparison of convergence metrics for the LIMO device with $N_p=4$ under a ventricular pressure $P_v=13.8$ kPa.} 
    (Left) Evolution of the maximum normalized vertical displacement $w/L$ as a function of actuation pressure. 
    (Middle) Convergence of the final maximum vertical deformation at $P_a=60$ kPa for different mesh densities with increasing number of degrees of freedom (Ndof). 
    (Right) Convergence of the Chamfer distance $d_\mathrm{c}$ between deformed surface point clouds for each mesh relative to the finest discretization (which has zero distance with itself).
    Maximum deformations are within 2\% between the medium and fine mesh.}
    \label{fig:limo_valid_metrics}
\end{figure}

%%%%%%%%%%%%%%%%%%%%%%%%%%%%%%%%
Next, we perform a mesh sensitivity study to assess the discretization errors of our LIMO soft artificial heart model.
We use $N_p=4$ and $P_v=13.8$ kPa as the reference case and perform static inflation simulations with three different mesh densities.
Following Section~\ref{appendix:solver_validation}, we employ quadratic quadrilateral shell elements (S8) with full integration, resulting in approximately 70k, 250k, and 400k degrees of freedom (Ndof) for the coarse, medium, and fine mesh, respectively.
Figures~\ref{fig:limo_valid_deformation} and \ref{fig:limo_valid_metrics} summarize the resulting deformation patterns and convergence metrics.

Figure~\ref{fig:limo_valid_deformation} compares the final deformed configurations for the three meshes and illustrates the associated cross-sectional evolution at relative height $y/L=0.5$ over the pressurization history (up to the peak actuation pressure). The coarse mesh under-resolves the edge buckling and therefore underpredicts the peak deformation relative to the two finer discretizations. The medium and fine meshes both capture the dominant buckling pattern and yield nearly indistinguishable deformation envelopes, suggesting mesh convergence of the global shape response.

To quantify this convergence, Figure~\ref{fig:limo_valid_metrics} reports (left) the maximum vertical deformation ($w/L$) against actuation pressure $P_a$, (center) the convergence of the final maximum vertical deformation at $P_a=60$ kPa, and (right) the Chamfer distance $d_c$ between deformed surface point clouds.
The Chamfer distance $d_c$ between two point clouds 
$P_1 = \{\vec{x}_i \in \mathbb{R}^3\}_{i=1}^n$ and 
$P_2 = \{\vec{x}_j \in \mathbb{R}^3\}_{j=1}^m$ is defined as the average distance between pairs of nearest neighbors between $P_1$ and $P_2$ i.e.
\begin{equation}
    d_\mathrm{c}(P_1, P_2) = \frac{1}{2n} \sum_{i=1}^n \Vert\vec{x}_i - \mathcal{NN}(\vec{x}_i, P_2)\Vert + \frac{1}{2m} \sum_{j=1}^m \Vert\vec{x}_j - \mathcal{NN}(\vec{x}_j, P_1)\Vert
\end{equation}
 and $\mathcal{NN}(\vec{x}, P)$ denotes the nearest neighbor of $\vec{x}$ in $P$,
 \begin{equation}
\mathcal{NN}(\vec{x}, P) = \underset{\vec{y} \in P}{\text{argmin}} \Vert\vec{x} - \vec{y}\Vert. 
\end{equation}

We find that maximum deformations differ by less than 2\% between the medium and fine mesh, and the the Chamfer distance satisfies $d_c$(medium, fine)$<5\times10^{-3}$.
On this basis, we use the medium mesh for all simulations reported in the main manuscript.

\clearpage
\setcounter{figure}{0} % reset figure counter
%%%%%%%%%%%%%%%%%%%%%%%%%%%%%%%%
\section{Deformation under various ventricular pressures and various pouch numbers}
\label{appendix:deformation}
\begin{figure}[H]
    \centering
    \includegraphics[width=\textwidth]{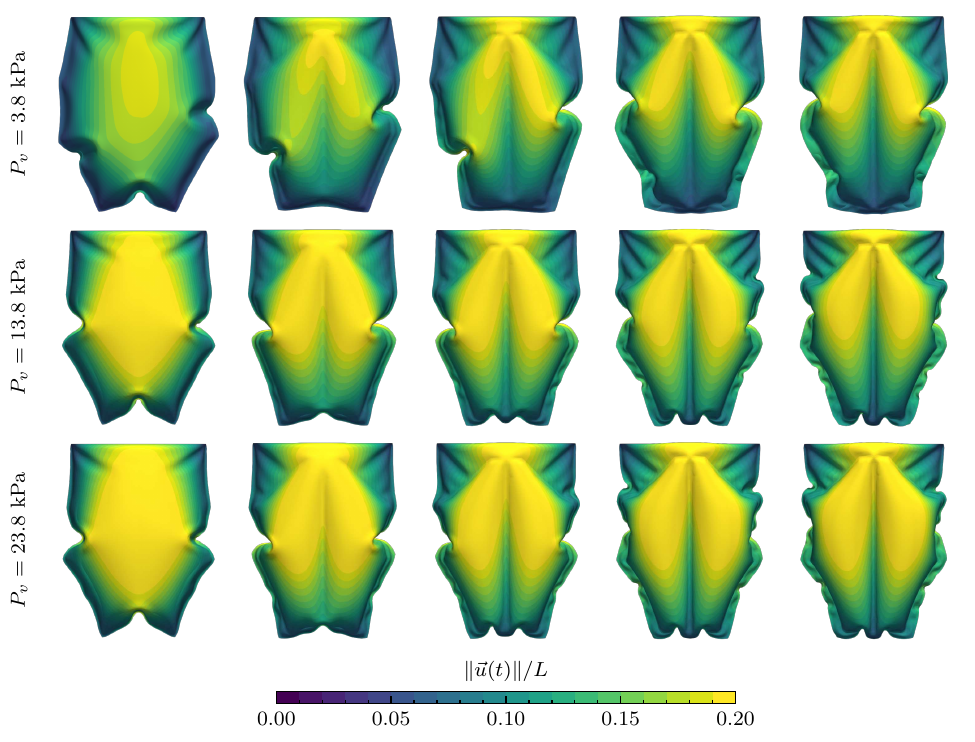}
    \caption{\textbf{Deformation of the device with $N_p=4$ under various ventricular pressures $P_v$.} Deformed geometry, colored with deformation magnitude, normalized by device length. Top, middle, and bottom rows correspond to ventricular pressures of 3.8, 13.8, and 23.8 kPa, respectively. The different snapshots are taken during the inflation of the pouches, starting with the left column at $P_a=0.0$ kPa and increasing the actuation pressure we move to the right. The actuation pressures in column 2, 3, 4 and 5 are 10$+P_v$, 20$+P_v$, 40$+P_v$, 60$+P_v$ kPa, respectively.}
    \label{fig:app_defprof_np4}
\end{figure}

\begin{figure}[H]
    \centering
    \includegraphics[width=\textwidth]{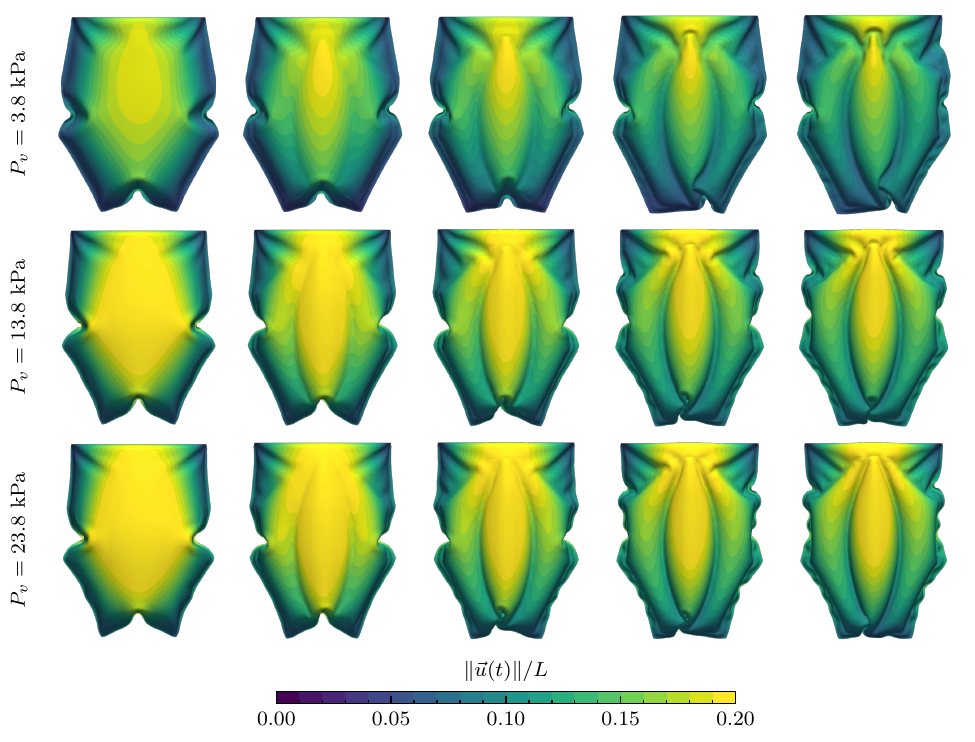}
    \caption{\textbf{Deformation of the device with $N_p=6$ under various ventricular pressures $P_v$.} Deformed geometry, colored with deformation magnitude, normalized by device length. Top, middle, and bottom rows correspond to ventricular pressures of 3.8, 13.8, and 23.8 kPa, respectively. The different snapshots are taken during the inflation of the pouches, starting with the left column at $P_a=0.0$ kPa and increasing the actuation pressure we move to the right. The actuation pressures in column 2, 3, 4 and 5 are 10$+P_v$, 20$+P_v$, 40$+P_v$, 60$+P_v$ kPa, respectively.}
    \label{fig:app_defprof_np6}
\end{figure}

\begin{figure}[H]
    \centering
    \includegraphics[width=\textwidth]{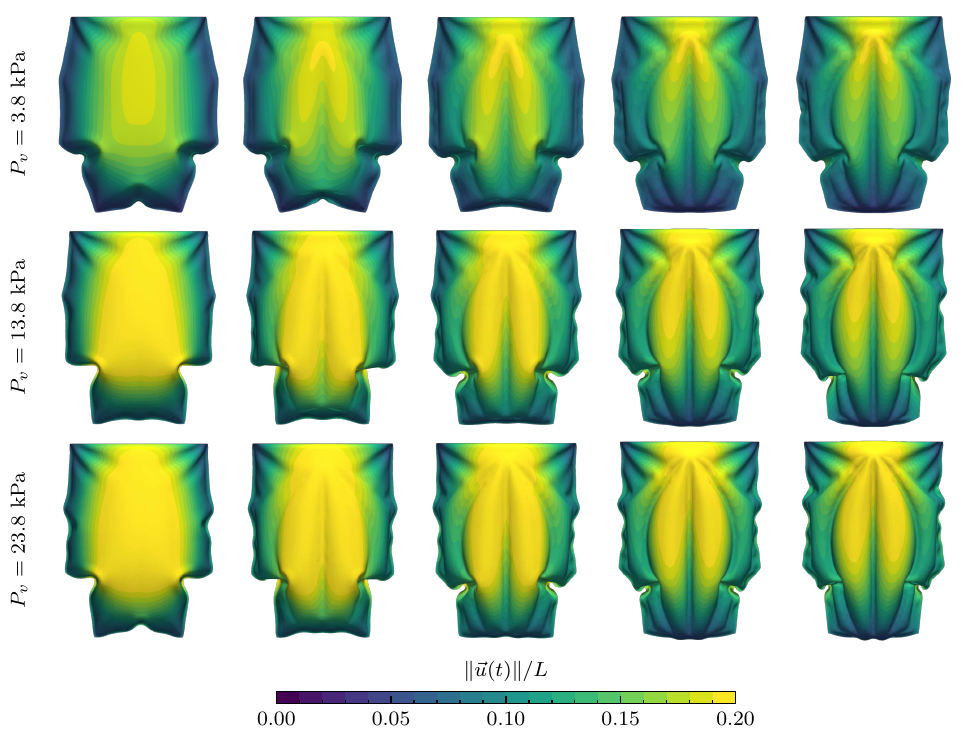}
    \caption{\textbf{Deformation of the device with $N_p=8$ under various ventricular pressures $P_v$.} Deformed geometry, colored with deformation magnitude, normalized by device length. Top, middle, and bottom rows correspond to ventricular pressures of 3.8, 13.8, and 23.8 kPa, respectively. The different snapshots are taken during the inflation of the pouches, starting with the left column at $P_a=0.0$ kPa and increasing the actuation pressure we move to the right. The actuation pressures in column 2, 3, 4 and 5 are 10$+P_v$, 20$+P_v$, 40$+P_v$, 60$+P_v$ kPa, respectively.}
    \label{fig:app_defprof_np8}
\end{figure}

\begin{figure}[H]
    \centering
    \includegraphics[width=\textwidth]{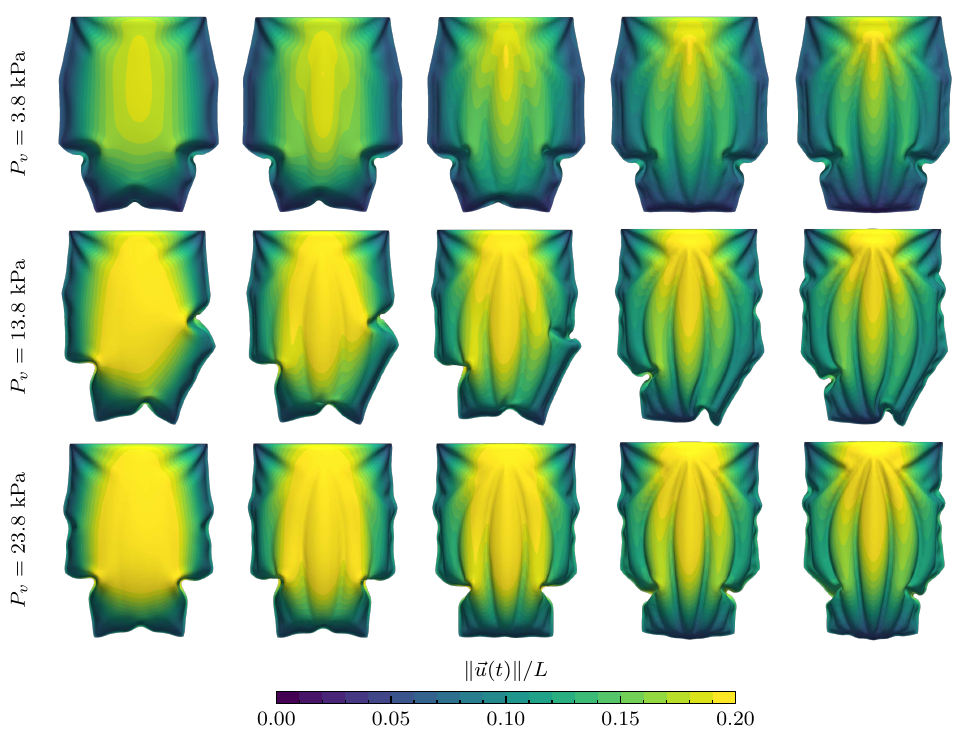}
    \caption{\textbf{Deformation of the device with $N_p=10$ under various ventricular pressures $P_v$.} Deformed geometry, colored with deformation magnitude, normalized by device length. Top, middle, and bottom rows correspond to ventricular pressures of 3.8, 13.8, and 23.8 kPa, respectively. The different snapshots are taken during the inflation of the pouches, starting with the left column at $P_a=0.0$ kPa and increasing the actuation pressure we move to the right. The actuation pressures in column 2, 3, 4 and 5 are 10$+P_v$, 20$+P_v$, 40$+P_v$, 60$+P_v$ kPa, respectively.}
    \label{fig:app_defprof_np10}
\end{figure}

%%%%%%%%%%%%%%%%%%%%%%%%%%%%%%%%
For completeness, we present the simulated time-varying deformation for the $N_p=4$ (Figure~\ref{fig:app_defprof_np4}) device, the $N_p=6$ (Figure~\ref{fig:app_defprof_np6}) device, the $N_p=8$ (Figure~\ref{fig:app_defprof_np8}) device, and the $N_p=10$ (Figure~\ref{fig:app_defprof_np10}) device respectively, starting at the end of the ventricular inflation $P_a=0$, until the end of the pressurization $P_a=P_{a,\text{max}}$.

%%%%%%%%%%%%
%% NP = 4
%%%%%%%%%%%%

%%%%%%%%%%%%
%% NP = 6
%%%%%%%%%%%%

%%%%%%%%%%%%
%% NP = 8
%%%%%%%%%%%%

%%%%%%%%%%%%
%% NP = 10
%%%%%%%%%%%%

\clearpage
\setcounter{figure}{0} % reset figure counter
%%%%%%%%%%%%%%%%%%%%%%%%%%%%%%%%
\section{Fatigue estimate for different ventricular pressures and pouch numbers}
\label{appendix:fatigue_all}
%%%%%%%%%%%%%%%%%%%%%%%%%%%%%%%%
We provide here the fatigue life estimate for all the different device geometries considered in this manuscript. For a detailed explanation of the method used to estimate the fatigue life, we refer the reader to Section~\ref{ssec:res_stress_fatigue}.

\begin{figure}[H]
    \centering
    \includegraphics[width=\textwidth]{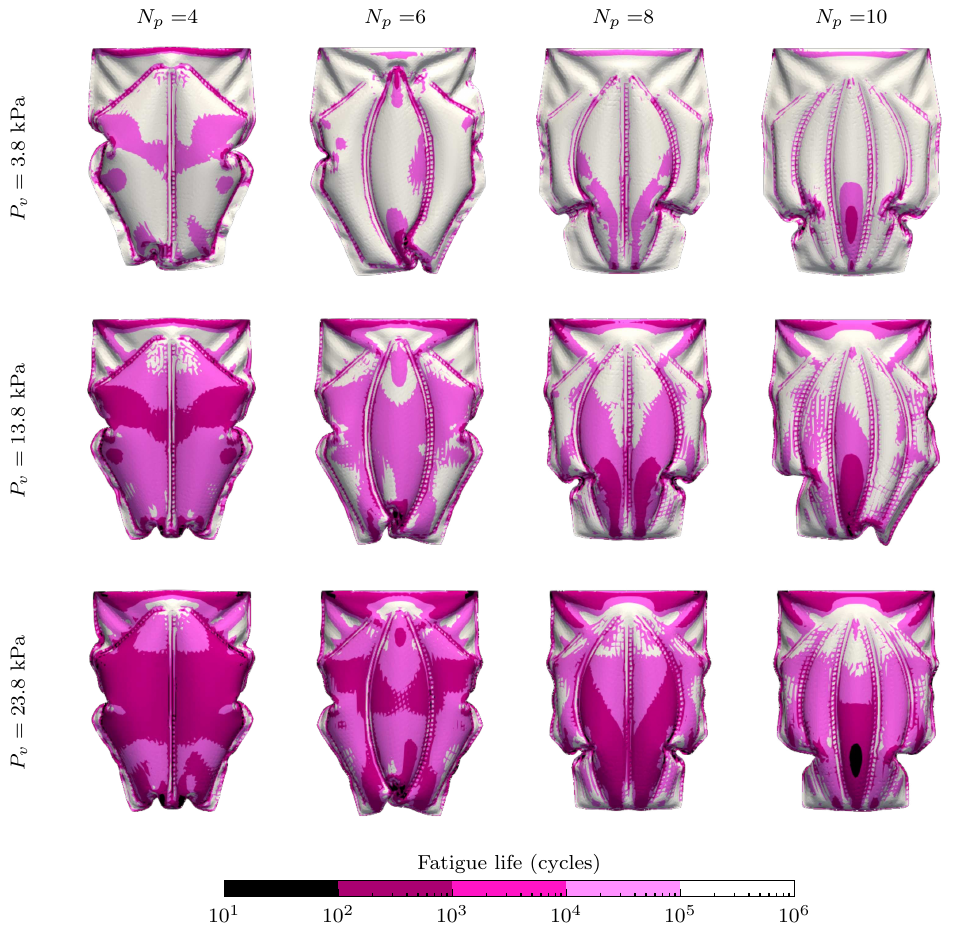}
    \caption{\textbf{Fatigue life estimate for $N_p \in \{4,6,8,10\}$ under various ventricular pressures $P_v$.} Isocontours of the expected number of cycles of a \textbf{pure} TPU device using the maximum principal strain in the membrane as a reference strain for the fatigue assessment see Section~\ref{ssec:res_stress_fatigue} and Figure~\ref{fig:fatigue_life} for a detail description of the method used.
    Note that we assume that the weakest points are near the seams on the TPU-TPU bonds, and thus we use a strain-cycle curve for pure TPU type only.
    The expected number of cycles is only valid in the seams region.
    Top, middle, and bottom rows correspond to ventricular pressures of 3.8, 13.8, and 23.8 kPa, respectively.}
    \label{fig:fatigue_all}
\end{figure}

\clearpage
\setcounter{figure}{0} % reset figure counter
%%%%%%%%%%%%%%%%%%%%%%%%%%%%%%%%
\newtext{\section{Sensitivity to initial conditions}
\label{appendix:biffurcation}
%%%%%%%%%%%%%%%%%%%%%%%%%%%%%%%%
During the reconfiguration of the device, non-linear buckling of the device's edges occur, see Section~\ref{ssec:res_stress_fatigue}. The location and symmetry of the resulting folds and buckled region is highly dependent on the initial conditions of the device, however, the number of folds is governed by the balance of internal and external energy, such that buckled regions can move, but their number should remain the same for different initial conditions.
We show this effect, and the resulting variation in pressure-volume relationship by generating an additional mesh, that also posses right/left symmetry on top of the original front/back symmetry.
The resulting deformation pattern during static inflation under a ventricular pressure $P_v=13.8$ kPa are shown in Figure~\ref{fig:symmetry_sensitivity}.}

\begin{figure}[H]
    \centering
    \includegraphics[width=\textwidth]{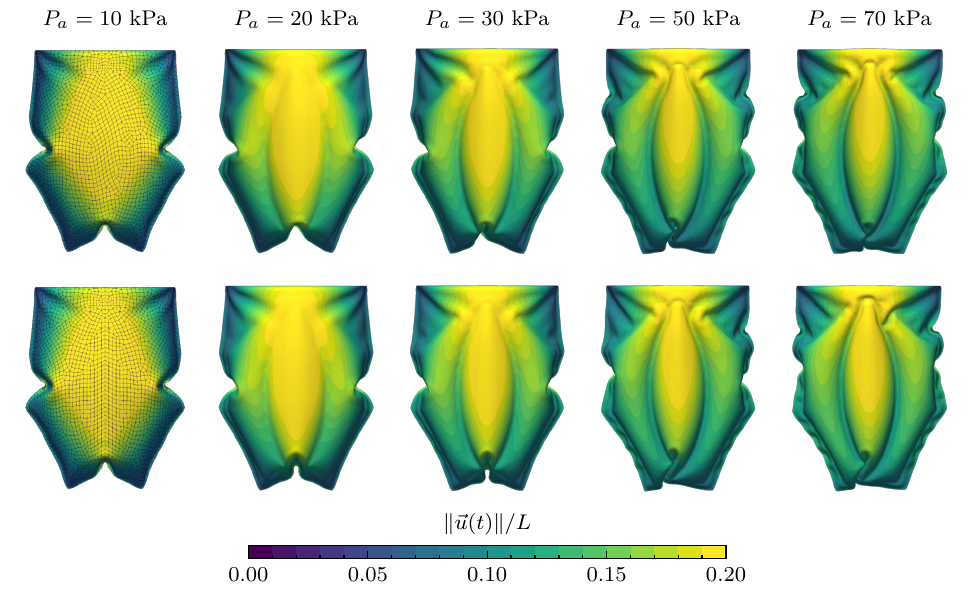}
    \caption{\textbf{Sensitivity of the device deformation for different initial conditions} Deformed geometry, colored with deformation magnitude, normalized by device length. Top, and bottom rows correspond a symmetric and double-symmetric mesh, respectively. The different snapshots are equally spaced during static inflation of a device with $N_p=6$ under a ventricular pressure $P_v=13.8$ kPa. The mesh density is shown on top of the results in the initial configuration $P_a=0.0$ kPa.}
    \label{fig:symmetry_sensitivity}
\end{figure}

\begin{figure}[H]
    \centering
    \includegraphics[width=\textwidth]{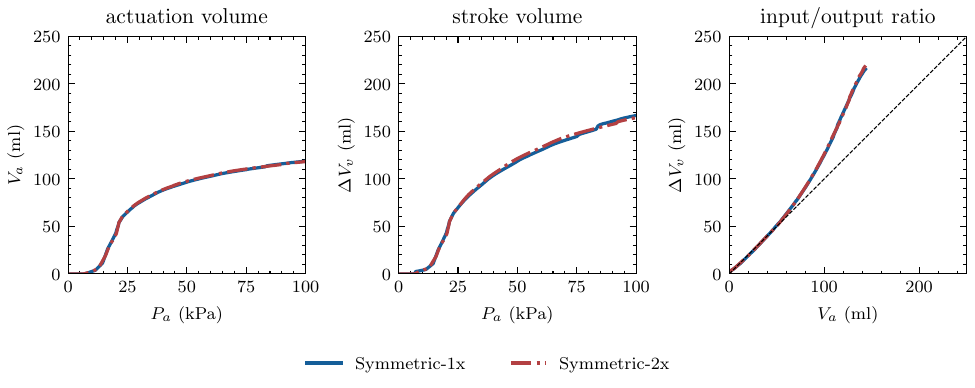}
    \caption{\textbf{Sensitivity of the pressure-volume relationship for different initial conditions.} Actuation volume against actuation pressure (left), stroke volume (center) and input-output ratio for two different meshes for a device with $N_p=6$ under a ventricular pressure $P_v=13.8$ kPa.}
    \label{fig:symmetry_pressures}
\end{figure}

\newtext{Pressure-volume relationship are very weakly affected by this different initial conditions and bifurcation of the solution, see Figure~\ref{fig:symmetry_pressures}. A similar observation is made if we alter the loading rate, the final solution show identical reconfiguration, but the position of the buckling/ folding can change.}

\clearpage
\bibliographystyle{elsarticle-num} 
\bibliography{library, references}

\end{document}